\documentclass[]{pasj02} 
\usepackage[switch,mathlines]{lineno}

\usepackage{physics}

\jyear{2024}
\Received{}
\Accepted{}

\begin{document} 

\title{ADF22-WEB: A giant barred spiral starburst galaxy in the $z=3.1$ SSA22 protocluster core}
\author{Hideki~Umehata$^{1,2}$}%
\author{C.C.~Steidel$^{3}$}
\author{Ian~Smail$^{4}$}
\author{A.M.~Swinbank$^{4}$}
\author{E.B.~Monson$^{5}$}
\author{David~Rosario$^{6}$}
\author{B.D.~Lehmer$^{7}$}
\author{Kouichiro~Nakanishi$^{8,9}$}
\author{Mariko~Kubo$^{10}$}
\author{Daisuke~Iono$^{8,9}$}
\author{D.M.~Alexander$^{4}$}
\author{Kotaro~Kohno$^{11,12}$}
\author{Yoichi~Tamura$^{2}$}
\author{R.J.~Ivison$^{13,14,15,16}$}
\author{Toshiki~Saito$^{17}$}
\author{Ikki~Mitsuhashi$^{18,8}$}
\author{Suo~Huang$^{8,2}$}
\author{Yuichi~Matsuda$^{8,9}$}

\altaffiltext{1}{Institute for Advanced Research, Nagoya University, Furocho, Chikusa, Nagoya 464-8602, Japan}
\altaffiltext{2}{Department of Physics, Graduate School of Science, Nagoya University, Furocho, Chikusa, Nagoya 464-8602, Japan}
\email{umehata@a.phys.nagoya-u.ac.jp}

\altaffiltext{3}{Cahill Center for Astronomy and Astrophysics, California Institute of Technology, MS249-17, Pasadena, CA91125, USA}

\altaffiltext{4}{Centre for Extragalactic Astronomy, Department of Physics, Durham University, South Road, Durham DH1 3LE, UK} 

\altaffiltext{5}{Department of Astronomy and Astrophysics, The Pennsylvania State University, 525 Davey Lab, University Park, PA 16802, USA}

\altaffiltext{6}{School of Mathematics, Statistics and Physics, Newcastle
University, Newcastle upon Tyne, NE1 7RU, UK}

\altaffiltext{7}{Department of Physics, University of Arkansas, 226 Physics Building, 825 West Dickson Street, Fayetteville, AR 72701, USA}

\altaffiltext{8}{National Astronomical Observatory of Japan, 2-21-1 Osawa, Mitaka, Tokyo 181-8588, Japan}

\altaffiltext{9}{Department of Astronomical Science, The Graduate University for Advanced Studies, SOKENDAI, 2-21-1 Osawa, Mitaka, Tokyo
181-8588, Japan}

\altaffiltext{10}{Astronomical Institute, Tohoku University, 6-3, Aramaki, Aoba, Sendai, Miyagi, 980-8578, Japan}

\altaffiltext{11}{Institute of Astronomy, Graduate School of Science, The University of Tokyo, 2-21-1 Osawa, Mitaka, Tokyo 181-0015, Japan}

\altaffiltext{12}{Research Center for the Early Universe, Graduate School of Science, The University of Tokyo, 7-3-1 Hongo, Bunkyo-ku, Tokyo 113-0033, Japan}

\altaffiltext{13}{European Southern Observatory (ESO), Karl-Schwarzschild-Strasse 2, D-85748 Garching, Germany}
\altaffiltext{14}{School of Cosmic Physics, Dublin Institute for Advanced Studies, 31 Fitzwilliam Place, Dublin D02 XF86, Ireland}
\altaffiltext{15}{
Institute for Astronomy, University of Edinburgh, Royal Observatory, Blackford Hill, Edinburgh EH9 3HJ, UK}
\altaffiltext{16}{ARC Centre of Excellence for All Sky Astrophysics in 3 Dimensions (ASTRO 3D)}

\altaffiltext{17}{Faculty of Global Interdisciplinary Science and Innovation, Shizuoka University, 836 Ohya,
Suruga-ku, Shizuoka 422-8529, Japan}

\altaffiltext{18}{Waseda Research Institute for Science and Engineering, Faculty of Science and Engineering, Waseda University, 3-4-1 Okubo, Shinjuku,
Tokyo 169-8555, Japan}


\maketitle

\begin{abstract}
In the present-day universe, the most massive galaxies are ellipticals located in the cores of galaxy clusters, harboring the heaviest super-massive black holes (SMBHs).
However the mechanisms that drive the early growth phase and subsequent transformation of these morphology and kinematics of galaxies remain elusive.
Here we report (sub)kiloparsec scale observations of stars, gas, and dust in ADF22.A1, a bright dusty starburst galaxy at $z=3.1$, hosting a heavily obscured active galactic nucleus and residing in a proto-cluster core. 
ADF22.A1 is a giant spiral galaxy with the kinematics of a rotating disk with rotation velocity $V_{\rm rot}=530\pm10$\,km\,s$^{-1}$ and diameter $>30$\,kpc.
The high specific stellar angular momentum of this system, $j_*=3400\pm600$\,kpc\,km\,s$^{-1}$, requires a mechanism to effectively spin-up ADF22.A1, 
indicating the importance of accretion from the cosmic web to supply both gas and angular momentum to galaxies in their early gas-rich starburst phase.
In its inner region, gas flows along dust lanes in a bar connected with the bright dusty core and the estimated mass ratio of a bulge to SMBH matches the local relation, suggesting that bars are a key mechanism to shape the early co-evolution of these components.
Comparison with cosmological simulations shows that ADF22.A1 will likely evolve into a massive elliptical at the present day, experiencing a significant reduction in angular momentum associated with subsequent galaxy mergers.
\end{abstract}

\pagewiselinenumbers   

\section{Introduction}

One of the fundamental correlations observed in galaxy evolution is the 
morphology-density relation. It is known that the cores of clusters in the local universe are populated by the oldest and most massive elliptical galaxies, while spiral galaxies are more common in the surrounding low-density environments (\cite{1980ApJ...236..351D}). 
In addition to their stellar populations, supermassive black holes (SMBHs) are known to be hosted by these galaxies, provides an important clue.
An important observational result that links the central SMBH with the host galaxy is a tight correlation between the mass of the central black hole, \(M_\mathrm{BH}\), and the mass of the host galaxy’s bulge (e.g., \cite{1998AJ....115.2285M}; \cite{2013ARA&A..51..511K}). This relation suggests a co-evolutionary pathway, where the growth of the SMBH and the host galaxy are regulated by common processes.
The connection between the environment and these galaxy populations showcases that environment play a key role in the formation and evolution of galaxies.

Recent observations and theoretical models suggest that the progenitors of massive ellipticals observed in the local Universe have formed most of their stars at high redshift ($z \gtrsim 2$) (\cite{2005ApJ...621..673T}; \cite{2006MNRAS.366..499D}).
Dusty star-forming galaxies (DSFGs), characterized by intense star formation and significant dust obscuration (for a review, \cite{2014PhR...541...45C}), are plausible progenitors (\cite{2014ApJ...782...68T};  \cite{2016ApJ...824...36C}).
DSFGs often harbor active galactic nuclei
(AGNs), which suggests that DSFGs also exhibit the growth
phase of SMBHs (e.g., \cite{2005ApJ...632..736A}).
In the last decades, several works reported the co-existence of DSFGs and proto-clusters at $z\sim3-4$ based on interferometric observations (e.g., \cite{2009ApJ...694.1517D};  
\cite{2015ApJ...815L...8U}; 
\cite{2016ApJ...828...56W}; \cite{2018ApJ...856...72O};
\cite{2018Natur.556..469M}).

The SSA22 proto-cluster at $z=3.09$ is one of the most remarkable overdensity at $z>2$ (\cite{1998ApJ...492..428S}; \cite{2004AJ....128.2073H}; \cite{2012AJ....143...79Y}) and offers one of the best targets to investigate the early formation of massive galaxies and the environmental dependence.
On top of single-dish surveys prior to the launch of the Atacama Large Millimeter/submillimeter Array (ALMA) (\cite{2004ApJ...611..725B}; \cite{2005MNRAS.363.1398G}; \cite{2005ApJ...622..772C}; \cite{2009Natur.459...61T}; \cite{2014MNRAS.440.3462U}), \citet{2015ApJ...815L...8U} (see also \citet{2019Sci...366...97U}) discovered that 16 DSFGs and 6 X-ray selected AGNs with $z_{\rm spec}\approx3.09$ are concentrated into the proto-cluster core, utilizing the contiguous 1mm mapping of the $2^{\prime\prime}\times3^{\prime\prime}$ core region.
This was the first discovery of the tight connection between such a large number of DSFGs/AGNs and proto-clusters in the early universe free from source confusion. The ALMA mosaic field is called as the ALMA Deep Field in SSA22 (ADF22) (\cite{2015ApJ...815L...8U}; \cite{2017ApJ...835...98U}).

ADF22.A1 at $z_{\rm spec}=3.09$ is the brightest DSFG in ADF22 situated within a large-scale Ly\,$\alpha$ filament extending over several Mpc (\cite{2019Sci...366...97U}). 
It is also the most luminous X-ray source in the field, hosting an intrinsically bright yet heavily obscured AGN (\cite{2010ApJ...724.1270T}; \cite{2023ApJ...951...15M}).
As such, ADF22.A1 offers a unique laboratory for exploring how the most massive galaxies and supermassive black holes (SMBHs) accumulate their mass, acquire their morphology, and undergo the transformations that ultimately evolve them into the most massive elliptical galaxies observed at $z=0$.
For years our understanding of the structure of the galaxy has been hampered in the past because of heavy dust extinction affecting its rest-frame UV appearance (\cite{2010ApJ...724.1270T}; \cite{2014MNRAS.440.3462U}).
However, with the advent of the James Webb Space Telescope (JWST) and ALMA, we can now resolve its structure and kinematics, providing unprecedented insights into the physical processes shaping the evolution of massive galaxies.

We adopt a standard concordance cosmology with $H_0=70$\,km\,s$^{-1}$,Mpc$^{-1}$, $\Omega_{\rm m}=0.30$, and $\Omega_\lambda=0.70$. Here $H_0$ is the Hubble constant. $\Omega_{\rm m}$ and $\Omega_\lambda$ are the matter density and dark energy density at the present time, respectively.
This gives a scale of 7.63\,kpc per arcsec at $z=3.09$.

\section{Observation and data reduction}

\begin{figure*}
\includegraphics[width=0.99\linewidth]{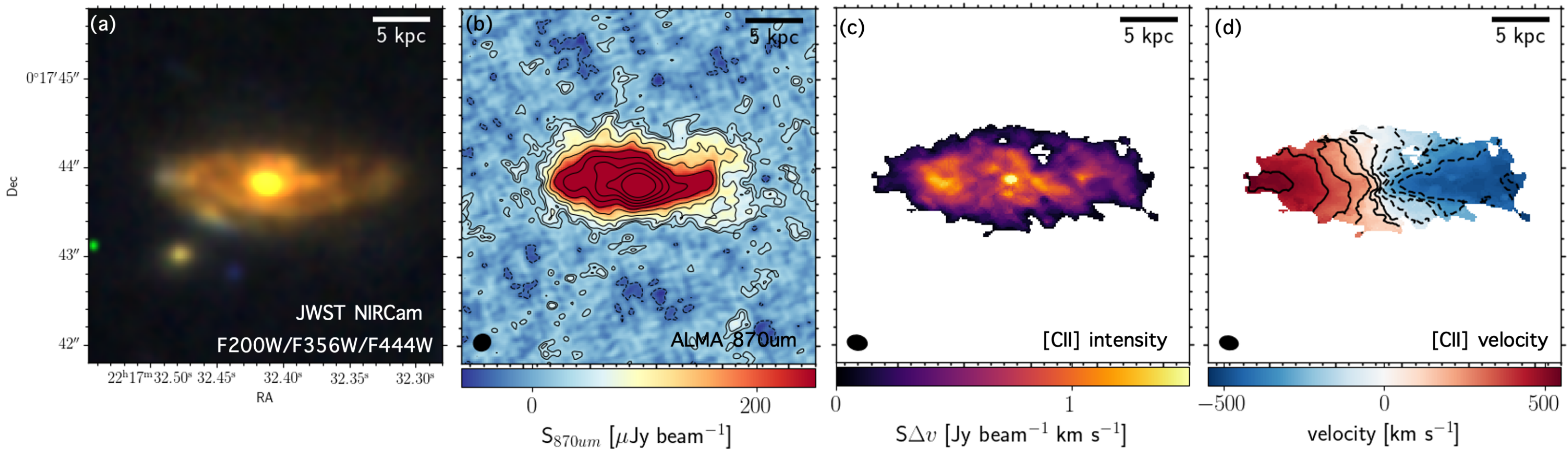}
    \caption{
    (a) A color image of a bright DSFG, ADF22.A1 at $z=3.09$ taken with JWST/NIRCam (blue: F200W, green: F356W, red: F444W), showing that ADF22.A1 resembles a giant spiral galaxy. 
    (b) The 870\,$\mu$m dust continuum image of ADF22.A1 taken with ALMA Band-7, which unveils active starburst activity throughout the disk.
    (c), (d) [C\,{\sc ii}] 158\,$\mu$m intensity and velocity maps of ADF22.A1 taken with ALMA Band-8, respectively. The intensity distribution shows a hint of spiral structure as also seen in stellar light. The velocity field is characteristic of a rotating disk, while non-circular motions are also observed as a twisted pattern at the center.
    The synthesized beam size for the ALMA data are shown in lower left corners of the relevant panels.
    }
    \label{fig:almamap}
\end{figure*}

\subsection{JWST NIRCam imaging}

The NIRCam imaging observation was performed on October 15 in 2023 as part of a JWST Cycle-2 program (PI Umehata, GO 3547) designed to cover ADF22, including ADF22.A1, with four filters (F115W, F200W, F356W, and F444W). The exposure time for each filter was 1869\,sec. We used the STANDARD subpixel dither pattern, while no primary dither was applied. We used the MEDIUM8 readout mode.
Data reduction was performed using  the JWST calibration pipeline (v.1.8.4), following the recipe provided by the CEERS team \citet{2023ApJ...946L..12B}. We applied ``snowball'' removal, wisp subtraction and ``1/f'' noise subtraction in addition to the standard reduction.
The 5$\sigma$ limiting magnitudes for a 0.1$^{\prime\prime}$ radius aperture are 
29.01, 29.15, 29.49, and 29.00 AB mag for F115W, F200W, F356W, and F444W, respectively. The point spread function (PSF) of each image was matched to that of the F444W image.
We refer to the joint ALMA-JWST effort within the proto-cluster core as ``ADF22-WEB''.

\subsection{ALMA observations}

\begin{table*}
\tbl{Summary of ALMA and JVLA observations}{%
\begin{tabular}{@{}lccccccc@{}}
\hline\noalign{\vskip3pt}
\multicolumn{1}{c}{Facility} & Band & Observable & Beam Size & PA (deg) & RMS (mJy) & Project ID \\[2pt]
\hline\noalign{\vskip3pt}
ALMA & Band8  & [C\,{\sc ii}] & $0.23^{\prime\prime} \times 0.17^{\prime\prime}$ & 78 & 0.47 mJy\,beam$^{-1}$ per 80\,km\,s$^{-1}$ & 2021.1.01406.S \\ 
 & Band8 & [C\,{\sc ii}] & $0.92^{\prime\prime} \times 0.84^{\prime\prime}$ & 74 & 0.8 mJy\,beam$^{-1}$ per 80\,km\,s$^{-1}$ & 2021.1.00041.S, 2021.1.01406.S \\ 
 & Band8  & 630\,$\mu$m & $0.078^{\prime\prime} \times 0.062^{\prime\prime}$ & -67 & 0.23\,mJy\,beam$^{-1}$ & 2022.1.00223.S\\
  & Band7  & 870\,$\mu$m & $0.21^{\prime\prime} \times 0.20^{\prime\prime}$ & 78 & 21\,$\mu$Jy\,beam$^{-1}$ & 2021.1.00071.S\\
  & Band6  & 1.1\,mm & $0.042^{\prime\prime} \times 0.042^{\prime\prime}$ & -4 & 12\,$\mu$Jy\,beam$^{-1}$ & 2019.1.00008.S\\
  & Band6  & 1.1\,mm & $0.079^{\prime\prime} \times 0.079^{\prime\prime}$ & -81 & 18\,$\mu$Jy\,beam$^{-1}$ & 2019.1.00008.S\\
JVLA  & Ka Band  & CO(1--0) & $3.23^{\prime\prime} \times 2.55^{\prime\prime}$ & -9 & 30-40\,$\mu$Jy\,beam$^{-1}$ per 100\,km\,s$^{-1}$ & 16A-357, 21A-346\\[2pt]
\hline
\end{tabular}}\label{table:observation_summary}
\end{table*}

The [C\,{\sc ii}] emission was observed using ALMA Band-8 as part of two ALMA observing programs.
The main program, to obtain a high angular resolution map of ADF22.A1, was carried out in April to June 2022 during the Cycle-8 project (PI Umehata, 2021.1.01406.S). The array configurations of C43-2 and C43-5 were utilized to achieve both high angular resolution as well as sensitivity to spatially extended emission. The on-source times were 11\,min and 39\,min, respectively. The precipitable water vapour (PWV) was typically 0.4\,mm.
Two spectral windows of 1.875 GHz bandwidth (with dual polarization) centered at 463.822 and 465.663\,GHz were employed.
Data reduction was performed using version 6.5.0. of the
Common Astronomy Software Applications (CASA) package. After continuum subtraction the data were mapped with the {\sc tclean} task, adopting Briggs weighting (robust = 0.5). This yielded a synthesized beam $0.23^{\prime\prime}\times0.17^{\prime\prime}$ at a position angle of 78 deg.
The typical noise level is $1\sigma=0.47$\,mJy\,beam$^{-1}$ per 80\,km\,s$^{-1}$ channel.
The parameters of the observations are summarized in Table \ref{table:observation_summary}, which also includes details of other observations described below.

As a part of the ALMA Cycle-8 project (PI, Umehata, 2021.1.00041.S), [C\,{\sc ii}] emission in ADF22.A1 was also observed in ALMA Band-8 in the C43-2 configuration in April 2022. The total on-source time was 18\,min, adopting the same correlator set-ups as the project above. Imaging used the
the {\sc tclean} task adopting natural weighting and {\sc uvtaper}=0.75$^{\prime\prime}$, to trace extended emission, was performed, combining all 29-minutes data taken with the C43-2 configuration in this project and the one above. The cube has a synthesized beam size of $0.92^{\prime\prime}\times0.84^{\prime\prime}$ with position angle 74\,deg.
The typical noise level is $1\sigma=0.8$\,mJy\,beam$^{-1}$ per 80\,km\,s$^{-1}$ channel.

As a part of the ALMA Cycle-9 project (PI Umehata, 2022.1.00223.S), dust continuum emission in ADF22.A1 was observed in ALMA Band~8. The C43-7 configuration was used in June 2023. The total on-source time was 64\,min. The image has representative frequency of 476.667\,GHz (630\,$\mu$m). 
The cube has a synthesized beam size of $0.078^{\prime\prime}\times0.062^{\prime\prime}$ with position angle -67\,deg.
The typical noise level is $1\sigma=0.23$\,mJy\,beam$^{-1}$. This image was used to obtain a high-resolution dust continuum map.

ALMA Band-7 observations to map the 870\,$\mu$m continuum were performed as a part of the Cycle-8 project (PI.Umehata, 2021.1.00071.S). The C43-3 and C43-6 array configurations were utilized in May and July in 2022, achieving on-source time of 8\,min and 21\,min, respectively. The correlator was set up with two spectral windows
of 1.875\,GHz bandwidth (dual polarization) each per sideband. The spectral windows had central frequencies of 336.5, 338.4, 348.5, and 350.5\,GHz, respectively.
Data reduction was performed using version {\sc casa} 6.5.0, mapped with the {\sc tclean} task, adopting natural weighting to yield a synthesized beam $0.21^{\prime\prime}\times0.20^{\prime\prime}$ with position angle 78 deg.
Typical noise level is $1\sigma=21$\,$\mu$Jy\,beam$^{-1}$.

Finally, ALMA Band-6 observations to map the 1.1\,mm dust continuum were carried out in August 2021 as a part of the Cycle-7 project (PI Umehata, 2019.1.00008.S). The array configuration was C43-8 and the total  integration time was 31\,min. The four spectral windows had central frequencies of 253.0, 254.8, 267.0, 269.0\,GHz.
Data reduction was performed using version 6.1.0. of the {\sc casa} package. 
The continuum image was created using the line-free channels with the the {\sc tclean} task. In the case of natural weighting, the resulting synthesized beam size is $0.042^{\prime\prime}\times0.042^{\prime\prime}$ with position angle $-4$ deg. The r.m.s. noise level is $1\sigma=12$\,$\mu$Jy\,beam$^{-1}$.
In the case of 
adopting natural weighting with {\sc uvtaper}=0.038$^{\prime\prime}$, the resulting synthesized beam size is $0.079^{\prime\prime}\times0.072^{\prime\prime}$ with position angle $-81$ deg. The r.m.s. noise level is $1\sigma=18$\,$\mu$Jy\,beam$^{-1}$. The tapered image is used in figures throughout the paper, while the original natural weighting image is utilized to measure the size of the dusty core.

\subsection{JVLA observation}

The CO(1--0) line at $z=3.09$ falls in the Ka-band of the Karl G. Jansky Very Large Array (JVLA). Observations were carried out over winter 2016 to summer 2021 in two programs (PI, Umehata, 16A-357, 21A-346).
We utilized the Wideband Interferometric Digital Architecture (WIDAR) correlator with 8-bit samplers. A continuous frequency range of $32.67-33.70$\,GHz and $27.67-28.70$\,GHz, which allowed us to observe the CO(1--0) line at the redshift range of $z=2.42-2.53$ and $z=3.02-3.17$ covering the $z=3.1$ proto-cluster.
The array configurations of C, CnB, and D were utilized to accumulate the
total on-source time was 24 hours.
The calibration was accomplished using the standard {\sc casa} pipeline for VLA (version 6.1.2.7).
All the $uv$-data were combined into a single data set. The data were mapped using {\sc tclean} in {\sc casa} with robust 2.0 weighting. 
The resulting synthesized beam size is $3.23^{\prime\prime}\times2.55^{\prime\prime}$ with position angle $-9$ deg. 
The achieved r.m.s. level per 100\,km\,s$^{-1}$ velocity channel is $30-40$\,$\mu$Jy beam$^{-1}$ at $28.0-28.4$\,GHz.

\subsection{Astrometry}

The astrometric accuracy of the ALMA images and cubes depends on (i) the S/N of the source and (ii) the quality of phase referencing and the positional uncertainty of the phase calibrator. First we measure the centroid of the phase calibrator J2226--0052 using the observed 870\,$\mu$m image, and confirm that the source position is matched within 1\,mas in the International Celestial Reference System (ICRS). Then we adopt the nominal positional accuracy for a compact source of 11\,mas, for the Band-7 data, which is derived by ${\rm pos}_{\rm acc}\approx {\rm beam}_{\rm FWHM}/{\rm SNR}/0.9$, 
where ${\rm beam}_{\rm FWHM}$ is the FWHM of the synthesized beam, SNR is the signal-to-noise ratio (up to 20), and 0.9 is a factor to account for a nominal 10\% signal decorrelation (ALMA technical Handbook and references therein). 
The JWST/NIRCam observation is not a mosaic but a single field observation. The limited spatial coverage of the NIRCam image means only a small number of Gaia stars are available. This makes it challenging to derive accurate absolute coordinates. Here our purpose is matching the JWST coordinate system with that of ALMA. 
We first align the four-band NIRCam images to each other, matching with Gaia stars.
Then we perform the emission profile fit for ALMA 870\,$\mu$m and NIRCam F444W images in a manner described in $\S 7$ also for other DSFGs in the field (H. Umehata et al. in preparation, for more details). We matched the ALMA and JWST images minimizing the systematic offsets between the peaks measured with the both images. Consequently both images match within a pixel scale of the JWST F444W image (0.063$^{\prime\prime}$\,pix$^{-1}$).

\section{Analysis and Results}

\subsection{Resolved views in stellar, gas, and dust components}

The JWST/NIRCam images trace the rest-frame optical-to-near-infrared wavelengths, have resolved the main stellar components at 0.2$^{\prime\prime}$ resolution (1.5\,kpc at $z=3.1$). Our images uncover spiral structure in the galaxy (Fig.~\ref{fig:almamap}a). 
The 870\,$\mu$m dust continuum 
 and emission in the fine-structure line of singly ionized carbon ([C\,{\sc ii}] 158\,$\mu$m) at kiloparsec resolution. These observations resolve the distribution and kinematics of the cold interstellar medium (ISM) in the young spiral galaxy.
As shown in Fig.~\ref{fig:almamap}b, the resolved dust continuum image uncovered that active star-formation activity, accompanied with significant dust production, occurs not only at the galactic center but across the whole disk.
The [C\,{\sc ii}] intensity distribution broadly traces the stellar light,  suggesting spiral arms (Fig.~\ref{fig:almamap}c, see also Fig.~\ref{fig:bar_summary}). The velocity field is characteristic of a regularly rotating disk in ADF22.A1, while non-circular motions cause a twisted pattern at the center (Fig.~\ref{fig:almamap}d).

\subsection{Panchromatic SED Fit}

\begin{figure}
\includegraphics[width=0.9\linewidth]{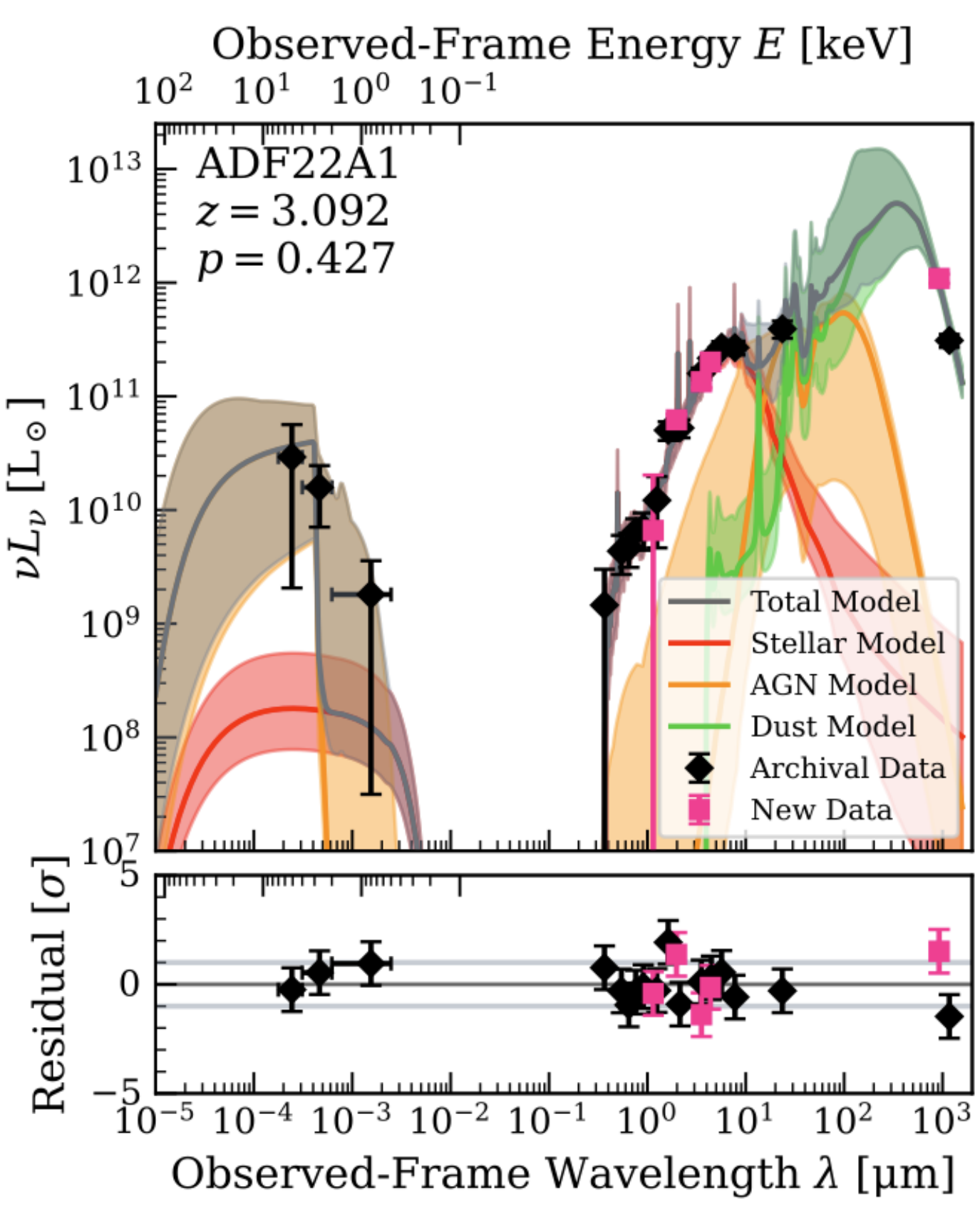}
    \caption{
    The best-fit SED model for ADF22.A1 from {\sc lightning} (\cite{2023ApJ...951...15M}). The shaded regions indicate the full range of the best-fitting 68\% of models.
    }
    \label{fig:lightning}
\end{figure}

Total flux densities of ADF22.A1 in the NIRCam images were measured with an elliptical aperture, which yields $<0.24$, $1.89\pm0.15$, $7.34\pm0.04$, $13.5\pm0.24$ $\mu$Jy for F115W, F200W, F356W, and F444W, respectively. We measure 870\,$\mu$m flux density $S_{\rm 870}=15.5\pm0.6$\,mJy based on a curve growth method. 
The X-ray-to-IR spectral energy distributions (SEDs) of ADF22.A1, except for these new measurements, were previously fit with the SED-fitting code {\sc lightning} (\cite{2023ApJ...951...15M}). This code allows us to simultaneously constrain emission from star-formation heated dust and AGN dust emission. We improve the fit by including the NIRCam and ALMA measurements.
The new best-fit SED model and the probability distribution function for key parameters are shown in Fig.~\ref{fig:lightning}. The fit yields SFR=$610_{-270}^{+340}$\,M$_\odot$\,yr$^{-1}$, log($M_*$/M$_\odot$)=11.4$^{+0.1}_{-0.2}$, log($M_{\rm BH}$/M$_\odot$)=8.5$^{+0.2}_{-0.3}$, and log $\lambda_{\rm Edd}$=$-0.9^{-0.4}_{-0.3}$, where $\lambda_{\rm Edd}$ is the Eddington ratio.
The estimates are generally consistent with the previous ones. 

\subsection{Profile Fit}

\begin{figure}
\includegraphics[width=0.99\linewidth]{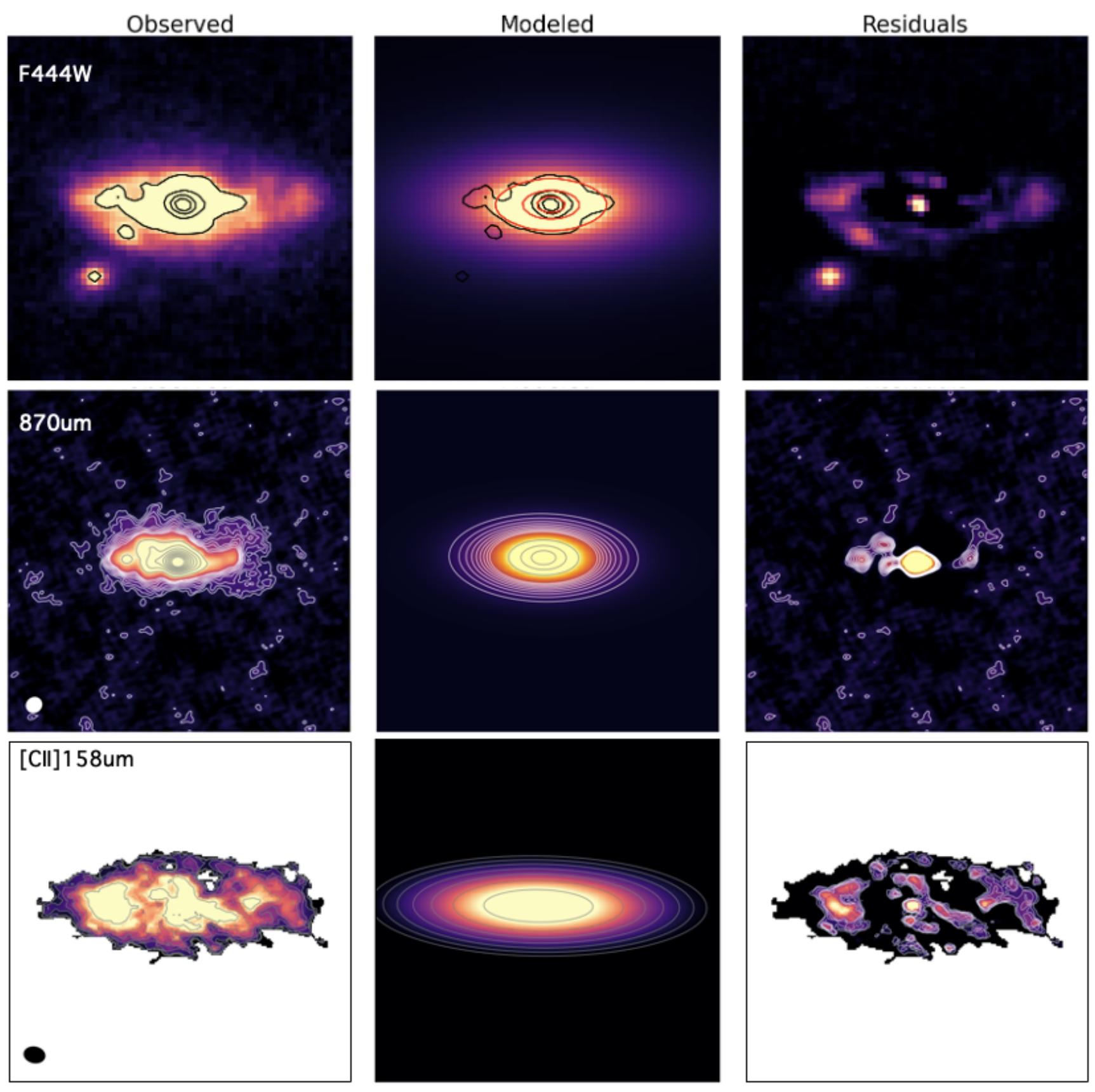}
    \caption{
    The results of profile fit of ADF22.A1. From top to bottom, cases of JWST/NIRCam F444W, ALMA 870\,$\mu$m, and 
    [C\,{\sc ii}] moment-0 maps are shown. In each row, observed maps are shown together with model and residual images. Each panel is 3.84$^{\prime\prime}$ in size.
    In the F444W maps, black contours show 20\%, 40\%, ..., 80\% of the peak flux in the observed or residual maps, while red contours show the case of the best-fit model.
    In the 870\,$\mu$m maps, white (gray) contours show $\sigma$, $3\sigma$, ... , $10\sigma$ ($20\sigma$, $30\sigma$, ... , $100\sigma$).
    In the [C\,{\sc ii}] maps, gray contours are $1.5\sigma$, $1.5\sigma^2$, ... , $1.5\sigma^7$, while white contours are from $1\sigma$ to $5\sigma$ in steps of $1\sigma$ (here $\sigma$ is for three Hanning-smoothed 40\,km\,s$^{-1}$ channels.
    As shown, the modeled S\'ersic profiles successfully reproduce the observed profile to a certain degree, while the 0.2$^{\prime\prime}$ maps leaves additional sub components, such as core and spiral arms.
    }
    \label{fig:galfit}
\end{figure}

\begin{figure}
\includegraphics[width=0.99\linewidth]{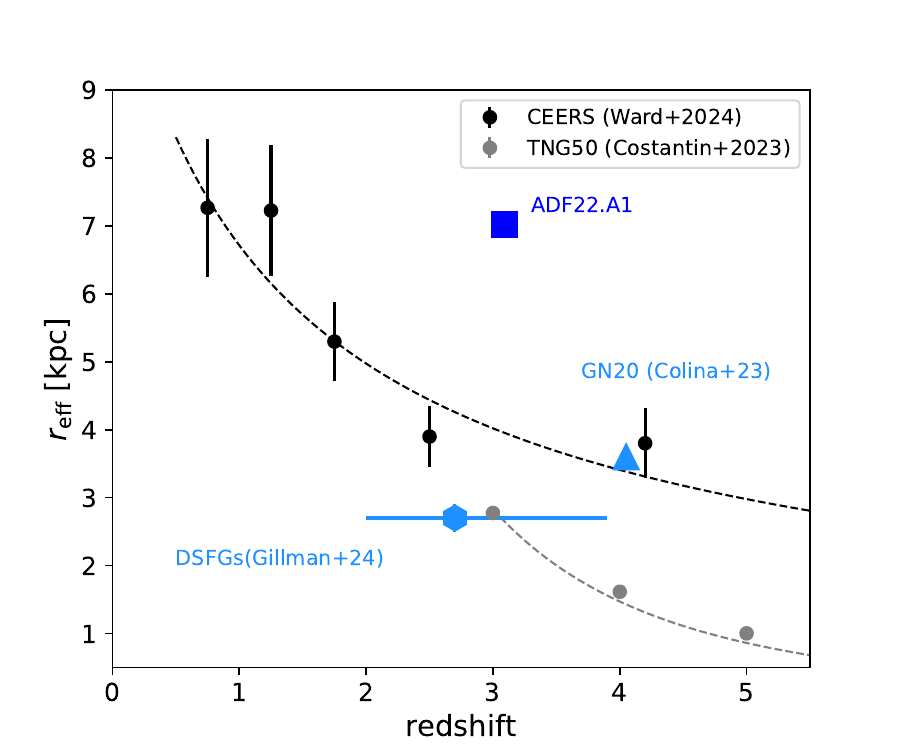}
    \caption{
    Effective radius of star-forming galaxies as a function of redshift. Expected trends for a stellar mass of $2\times10^{11}$\,M$_\odot$ of the size--mass relations for star-forming galaxies are illustrated for JWST observations (\cite{2024ApJ...962..176W}) and simulations (\cite{2023ApJ...946...71C}). The effective radius of ADF22.A1 is larger than other star-forming galaxies, including typical $z\sim3$ DSFGs, and equivalent to $z\sim1$ galaxies, showcasing an accelerated size and mass growth in the $z=3.1$ proto-cluster core.
    }
    \label{fig:z_size}
\end{figure}

We performed S\'ersic profile fitting using the F444W image using {\sc galfit} (\cite{2002AJ....124..266P}; \yearcite{2010AJ....139.2097P}). The image traces the rest-frame near-infrared emission around 1\,$\mu$m, which is expected to be sensitive to stellar mass distribution.
The best-fit model and residula images are shown in Fig.~\ref{fig:galfit}.
This best-fit model has an effective radius $r_{\rm e, F444w}=7.0\pm0.1$\,kpc and S\'ersic index $n=1.7\pm0.1$.

We compare the measured size of the stellar emission with other works based on JWST NIRcam observation and simulations in Fig.~\ref{fig:z_size}. It has been known that there is a trend between the galaxy size and stellar mass (\cite{2014ApJ...788...28V}). For a fair comparison, we here focus on works which uses the rest-frame optical-to-near infrared wavelengths proved with JWST images (or simulations for JWST). The size-mass relation derived based on the NIRCam images taken with the CEERS survey was measured  (\cite{2024ApJ...962..176W}), which provides a suitable comparison sample (note that most of their samples are based on photometric redshifts). We plot the derived effective radius for each redshift bin, adopting $M_*=2\times10^{11}$\,M$_\odot$, which is equivalent to the stellar mass of ADF22.A1. We note that we use the relation for star-forming galaxies in this work, while they discuss both star-forming galaxies and passive galaxies separately. 

The result of mock observations of TNG50 simulations are also shown, adopting the same stellar mass (\cite{2023ApJ...946...71C}). The best-fit functions for the observation and simulation are $r_{\rm eff}/{\rm kpc}=11.2\times(1+z)^{-0.74}$ and $r_{\rm eff}/{\rm kpc}=165.4\times(1+z)^{-2.9}$, respectively. Note that there are several caveats. The sampling of the CEERS survey data is sparse at $z>3$, resulting in a single redshift bin to cover $z=3-5.5$. Meanwhile, the simulation models galaxies at $z=3-6$ only. Galaxies with $M_*\gtrsim2\times10^{11}$\,M$_\odot$ are quite rare at $z\gtrsim3$ in both observations and simulations, and the size-mass relation in the mass range largely relies on extrapolation of the lower-mass range.  Nevertheless, they provide useful comparison at $z\simeq3$.
As shown, the effective radius of ADF22.A1 at $z=3.09$ is about $(1.8-2.5)\times$ larger than the representative values expected at the redshift from observations or simulations. The effective radius is larger than any galaxies in the CEERS survey at $z>3$ (\cite{2024ApJ...962..176W}), which demonstrate that ADF22.A1 is one of the largest galaxies at such an early epoch. The effective radius is equivalent to that of typical star-forming galaxies at $z=0.5-1.0$, suggesting accelerated size growth of a stellar disk in the violently growth phase in a rare $z=3.1$ proto-cluster core with a space density below those sampled by recent simulations. The S\'ersic index is significantly larger than unity, giving further evidence for the existence of a growing bulge. 

In Fig.~\ref{fig:z_size}, we also compared the result with other measurements on stellar sizes of DSFGs based on JWST/NIRCam. Eighty DSFGs observed by the PRIMER project at $z=2.7_{-0.7}^{+1.2}$ have effective radius $r_{\rm e, F444W}=2.7\pm0.2$\,kpc (\cite{2024arXiv240603544G}) with two relatively large examples with $r_{\rm e, F444W}\sim7$\,kpc. GN20 at $z=4.05$, which is a member of the proto-cluster at the field, have an effective radius $r_{\rm e, F560W}=3.6\pm0.03$\,kpc (\cite{2023A&A...673L...6C}). 
It is also reported that A1489-850.1 at $z=4.26$ has $r_{\rm e, F444W}=3.8\pm0.4$\,kpc (\cite{2023ApJ...958...36S}).
While these are relatively larger than other DSFGs at a lower redshift, their sizes are still only about a half of ADF22.A1. Thus the large size of ADF22.A1 is unusual among the DSFG population.

We also model the 0.2$^{\prime\prime}$ 870\,$\mu$m dust continuum emission and [C\,{\sc ii}] total flux with a S\'ersic profile (Fig.~\ref{fig:galfit}). ADF22.A1 has the peak flux density $S_{\rm 870}=2.61 \pm0.02$\,mJy\,beam$^{-1}$, which corresponds to the signal-to-noise ratio (SNR) $\approx124$. We masked $5\sigma$ peaks as presented in \citet{2019ApJ...876..130H} in the case of 870\,$\mu$m dust continuum, which yielded the best-fit model with an effective radius $r_{\rm e, 870}=4.7\pm0.1$\,kpc and S\'ersic index $n=0.9\pm0.1$. 
For [C\,{\sc ii}], the $^{\rm 3D}${\sc barolo} package (\cite{2015MNRAS.451.3021D}) was used to extract the [C\,{\sc ii}] emission from the cube with 40\,km\,s$^{-1}$ bins to make the moment-0 map. The genuine emission was identified with the {\sc search} task with {\sc SNRCUT}=4.5, {\sc GROWTHCUT}=4.5, and {\sc MINCHANNELS}=3. The [C\,{\sc ii}] peak flux is $1.95\pm0.07$ Jy\,beam$^{-1}$\,km\,s$^{-1}$ (SNR$\approx28$, by assuming that three channels as the typical channel width to combine).
We fit a S\'ersic profile without a bright emission mask and obtained $r_{\rm e, [CII]}=7.6\pm0.2$\,kpc and S\'ersic index $n=0.4\pm0.1$.
%
%
We note that the models still leave residuals, which demonstrate the power of high-resolution images at the 0.2$^{\prime\prime}$ resolution. These substructures include a core, clumps, and spiral arms. The spiral arms are indicated by all three components (stars, dust, and gas) while the relative strength among the three tracers show significant variation. 

\subsection{Kinematic modeling}

\begin{figure*}
\includegraphics[width=0.95\linewidth]{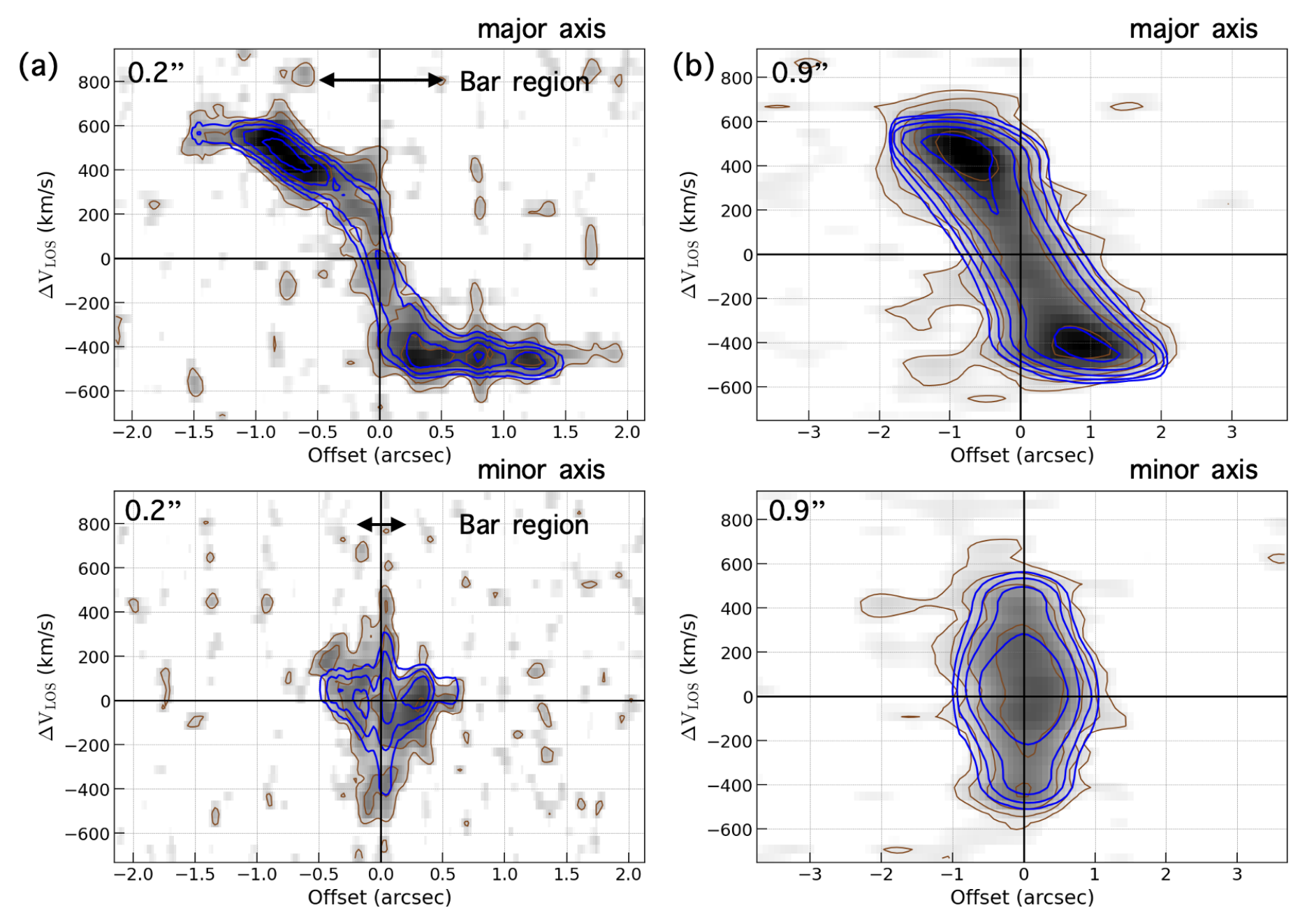}
    \caption{
    Position velocity diagrams of the [C\,{\sc ii}] emission at 0.2$^{\prime\prime}$ and 0.9$^{\prime\prime}$ along the major and minor axes as labelled. The background image and brown contours ($2\sigma$,$4\sigma$,$8\sigma$,$16\sigma$, and $32\sigma$) represent the observed emission. The best-fit 3D model derived with the {\sc 3d barolo} is overlayed in blue contours. The model globally reproduces the observed emission for the two cases with different angular resolutions. There are some discrepancy in the innermost region (denoted as ``bar region'') in the 0.2$^{\prime\prime}$ case, which encompasses the ``twisted'' pattern in the velocity field shown in Fig.~\ref{fig:almamap}. The innermost velocity structure is further investigated in \S \ref{dis:bar}.
    }
    \label{fig:barolo}
\end{figure*}

We made [C\,{\sc ii}] data cubes and fitted them with a rotating disk model to derive the kinematic properties. 
As presented in Fig.~\ref{fig:almamap}, the observed [C\,{\sc ii}] velocity field shows a generally monotonic velocity gradient, indicating a rotating gas disk in the spiral galaxy. A twisted pattern at the center also suggests the presence of non-circular motion (see also Fig.~\ref{fig:bar_summary}b for the detailed view). 
The $^{\rm 3D}${\sc barolo} package was utilized to model the [C\,{\sc ii}] kinematics as a rotating disk, which also allows a chance to isolate the non-circular components. In addition to the original 0.2$^{\prime\prime}$ resolution cube, we used a smoothed cube with a 0.9$^{\prime\prime}$ resolution, which improves the detectability of emission in the outskirts and suppresses the deviations produced by the non-circular motions at the center.

First we made use of the high-resolution cube (0.2$^{\prime\prime}$). The cube was created with a 40\,km\,s$^{-1}$ bin, which was then Hanning-smoothed with the {\rm smoothspec} function. We adopted a disk model where the kinematic center ($x_0$, $y_0$), position angle (PA), inclination ($i$), systemic velocity ($V_{\rm sys}$), rotation velocity ($V_{\rm rot}$), and velocity dispersion ($\sigma$) were free-parameters. The scale-height of the disc was fixed to be 300\,pc (\cite{2021Sci...371..713L}). 
We adopted a ring width of $0.097^{\prime\prime}$, which is half of the resolution element $\sqrt{\theta_x + \theta_y}$, where $\theta_{\rm x}$ and $\theta_{\rm y}$ are the major and minor axes of the synthesized beam. Considering the complex gas morphology (Fig.~\ref{fig:almamap}), a disk model with 
non-axisymmetric gas distributions was studied by renormalizing the flux density of each spatial pixel to the observed intensity map.
The evaluated inclination generally ranges from $i=70$\,deg to $i=75$\,deg with a typical errors of 4\,deg. The median value is $i=73$\,deg. For position angle, PA=$87-96$\,deg (a median value PA=92\,deg with errors of 5\,deg) are estimated. The kinematic center position matches with peaks of F444W and 870\,$\mu$m dust continuum emissions within the F444W pixel sampling.
 Fig.~\ref{fig:barolo}a shows the position-velocity (PV) diagram of the best-fit model, compared to the observed emission. While the model successfully reproduces the observed emission as a whole, there are some discrepancy in the inner most region comprising the twisted pattern. Errors on the derived parameters become large, reflecting the influence of the non-circular motion at the center. 
An elliptical region, which is centered at the kinematic center and aligned along the kinematic major axis, is defined to isolate the innermost region. This region is denoted as the ``bar region'' since the region corresponds to the area of the dusty core and offset ridges (Fig.~\ref{fig:bar_summary}). We will discuss this aspect more in \S \ref{dis:bar} later, together with possible influence of AGN.

The 0.9$^{\prime\prime}$ resolution cube is used to obtain the overall trend of the velocity structure, which is insensitive to the innermost non-circular motion and sensitive to the extended emission in the outer region. We adopt and fix the obtained median values of the parameters during the run for the 0.2$^{\prime\prime}$ cube except for rotation velocity and velocity dispersion. We adopt a ring width of 0.$^{\prime\prime}$35, which is 40\% of the resolution element, considering the consistency with spatial resolution and the accuracy of velocity field estimation. In the fit, rotation velocity and velocity dispersion are estimated. The PV diagram of the modeled and observed emissions are shown in Fig.~\ref{fig:barolo}. In this case, the model generally reproduces the observed emission.

It is suggested that the velocity dispersion derived by $^{\rm 3D}${\sc barolo} in the standard way can be significantly underestimated when the velocity resolution is not sufficient, which may be a typical situation in observations of galaxies at high redshift (\cite{2023A&A...672A.106L}). We 
use the task {\sc spacepar} in
$^{\rm 3D}${\sc barolo} to investigate a global minimum in the $V_{\rm rot}$-$\sigma$ space and obtain a better constraint on velocity dispersion, fixing other parameters. The explored parameter space is a range of $V_{\rm rot}$ from 1 to 700\,km\,s$^{-1}$ and $\sigma$ from 1 to 100\,km\,s$^{-1}$, respectively. Steps are 2\,km\,s$^{-1}$ for each. The minimum $\sigma$ is in a range from 44 to 60\,km\,s$^{-1}$ at $r=0.63^{\prime\prime}$ to 1.02$^{\prime\prime}$ (4.8 to 7.8\,kpc). The area corresponds to bright spiral arms and disks away from the central bulge and bar regions. This is close to the velocity resolution (80\,km\,s$^{-1}$) and hence we perform the similar test for a cube with 20\,km\,s$^{-1}$ bins (the velocity resolution after Hanning smoothing is 40\,km\,s$^{-1}$). This test resulted in a similar $\sigma$ range, which support the validity of the measurement with the original cube. We conservatively adopt a representative $\sigma=60$\,km\,s$^{-1}$.  

\begin{figure*}
\includegraphics[width=0.99\linewidth]{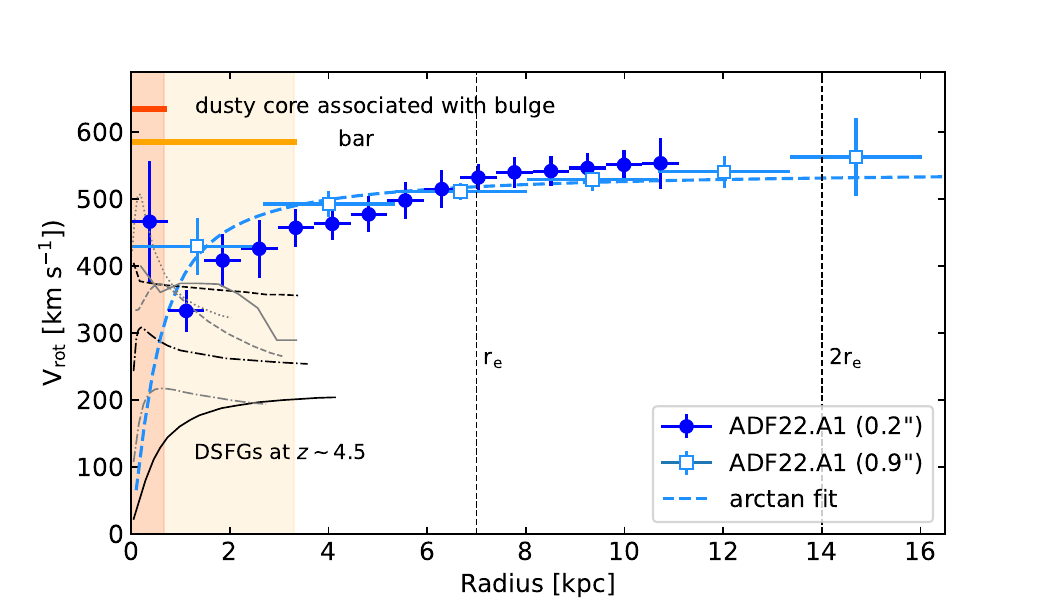}
    \caption{
The observed rotation curve of ADF22.A1 at $z=3.09$ derived from [C\,{\sc ii}] cubes at 0.2$^{\prime\prime}$ (blue circles) and 0.9$^{\prime\prime}$ (cyan squares), respectively. 
The smoothed data is beneficial to trace the overall trend. As the best-fit arctangent model is overlaid (cyan dashed line), the gas disk of ADF22.A1 is generally characterized as a rotating disk with a flat rotation curve $V_{\rm rot}=530$\,km\,s$^{-1}$ (at $r=2{\rm r}_{\rm e, F444W}$, where ${\rm r}_{\rm e, F444W}$ is the stellar effective radius measured with the F444W image. 
The high resolution data shows deviation from the model curve in the inner region, indicating the influence of potentials associated with the bar and bulge (as the sizes of them are indicated).
Other well-resolved rotation curves based on [C\,{\sc ii}] of $z\sim4.5$ DSFGs are also plotted (\cite{2020Natur.584..201R}; \yearcite{2021MNRAS.507.3952R}; \cite{2021Sci...371..713L})). The comparison between ADF22.A1 and such typical DSFGs highlights the unusually large extent and $V_{\rm rot}$ of ADF22.A1.
}
    \label{fig:rotationcurve}
\end{figure*}

The evaluated rotation velocity is plotted as a function of radius in Fig.~\ref{fig:rotationcurve} for both cases. We fit the rotation curve based on the 0.9$^{\prime\prime}$ cube with an arctangent model:

\begin{equation}
v(r) = \frac{2}{\pi} v_{\infty} \arctan\left(\frac{r}{r_t}\right)
\end{equation}

where $v_{\infty}$ is the asymptotic velocity at sufficiently large radii and ${r_t}$ is the scale radius at which the rotation curve transitions from rising to flat. 

The best fit model is also shown in Fig.~\ref{fig:rotationcurve} with $v_{\infty}=540$\,km\,s$^{-1}$ and ${r_t}=0.52$\,kpc. The rotation curve from the 0.2$^{\prime\prime}$ cube is generally consistent with that of the 0.9$^{\prime\prime}$ cube. Potential deviations are seen in the inner region ($r\lesssim4$\,kpc). The steep rise in the innermost bin can be caused by the compact gravitational potential associated with the stellar bulge, as other works report in some DSFGs (\cite{2021Sci...371..713L}). 
The derived rotation velocity $V_{\rm rot}$ remains flat with $V_{\rm rot}=530\approx$\,km\,s$^{-1}$ in the outer disk, extending to $r\approx15$\,kpc. 
We measure a rotation velocity $V_{\rm rot}$ using the arctangent model. At a  radius of $r=2{\rm r}_{\rm e}$, $V_{\rm rot}=530\pm10$\,km\,s$^{-1}$.
Adopting this value as a representative $V_{\rm rot}$, the ratio between the rotation velocity and velocity dispersion is $V_{\rm rot}/\sigma=8.8\pm1.5$. Hence rotation support dominates pressure support in ADF22.A1. Based on this result, we assume that the rotation velocity is approximated equal to the circular velocity.

The nature of ADF22.A1, a huge and rapidly-rotating disk, is distinguished among any known galaxies in the early universe.
The rotation velocity and extent of the gas disk are several times larger than those reported for DSFGs at $z\sim4-5$ (\cite{2020Natur.584..201R}; \yearcite{2021MNRAS.507.3952R}; \cite{2021Sci...371..713L})
(Fig.~\ref{fig:rotationcurve}), and also uniquely large among DSFGs at $z\sim2$\cite{2023arXiv231208959A}.
One possibly resemble case is GN20 at $z=4.05$ (\cite{2012ApJ...760...11H}). While the CO(2--1) data obtained with the VLA does not have sufficient quality for fair comparison with ADF22.A1, the reported maximum rotational velocity of $v_{\rm max} = 575 \pm 100$\,km\,s$^{-1}$ is comparable. They reported the emission extends to $r=7$\,kpc, which is about a half of that of ADF22.A1. Interestingly GN20 also resides in a proto-cluster at $z=4$ similar to ADF22.A1. 
There are also intriguing discovery about normal star-forming galaxies (not DSFGs). Large disk galaxies at $z\sim3$ in an overdensity are recently reported (\cite{2023ApJ...942L...1W}; \cite{2024arXiv240917956W}). While the rotation velocity is not so large ($v_{\rm rot} \sim 200-300$\,km\,s$^{-1}$), the disk size is similar to ADF22.A1.
There may be a common driver to form rapidly rotating disks and/or giant disks in a proto-cluster environment. As we discussed in the paper, both cold gas accretion and gas-rich major-mergers are expected to work efficiently in such an environment.

A difference between rotation velocities based on the two resolution cubes is identified in the inner region ($r\lesssim4$\,kpc),
while both measurements are consistent in the outer part. A steep rise in $V_{\rm rot}$ in the innermost region may be caused by a bulge potential (\cite{2021Sci...371..713L}), while the adjacent decline indicates the effect of a bar flow (\cite{2009PASJ...61..441H}).

\subsection{Resolved stellar profile and specific stellar angular momentum}

\begin{figure*}
\includegraphics[width=0.95\linewidth]{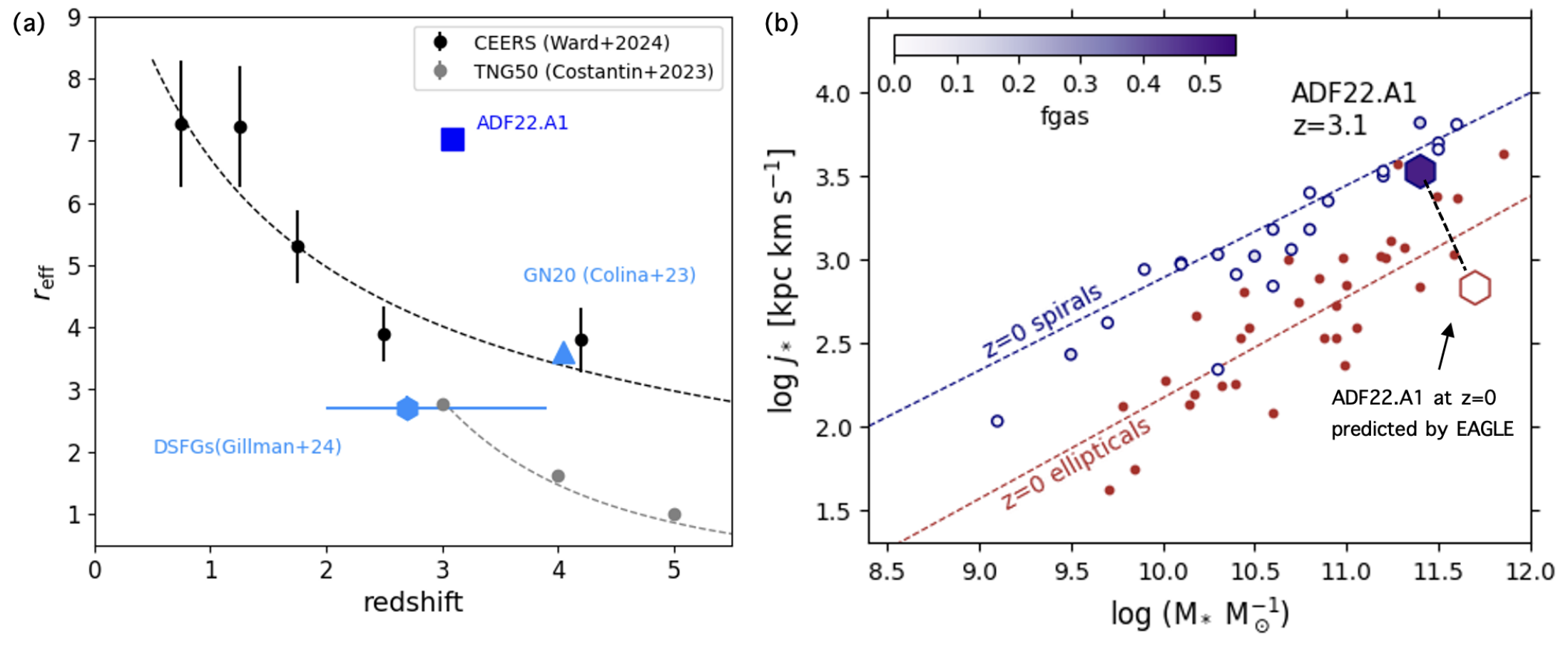}
    \caption{
    (a) Profile of stellar mass surface density. The best-fit model of bulge, disk, and the combined profiles are also shown. The excess in the innermost region reflects the stellar bulge, while the bar may also contributes to the excess. Since the current photometry is not sensitive enough to measure the profile at outer parts, we extrapolate it, assuming an exponential disk. 
    (b) cumulative fraction of the stellar mass $M_*$, rotation velocity $V_{\rm rot}$, and specific stellar angular momentum $j*$ as a function of radius for ADF22.A1. As shown, specific stellar angular momentum, calculated using stellar mass distribution and rotation curves, converges at a large radius. 
    }
    \label{fig:sAM}
\end{figure*}



The stellar specific angular momentum ($j_*=J_*/M_*$, where $J_*$ is total stellar angular momentum) involves three uncorrelated variables; a length scale, and a rotation-velocity, is one of the most fundamental quantities to describe kinematics and structures of galaxies (e.g., \cite{1980MNRAS.193..189F}; \cite{2012ApJS..203...17R}). The specific stellar angular momentum is given by

\begin{equation}
j_* = \frac{J_*}{M_*} = \frac{\int_{r_*} (\mbox{\boldmath $r$} \times \bar{\mbox{\boldmath $v$}}) \rho_* d^3 \mbox{\boldmath $r$}}{\int_{r} \rho_* d^3 \mbox{\boldmath $r$}}
\end{equation}
where $\vb*{r}$, $\bar{\vb*{v}}(\vb*{r})$, and $\rho(\vb*{r})$ are the position,  mean-velocity vectors, and the three-dimensional density of the stars.
This equation (2) can be reduced to a one-dimensional integral and given by:
\begin{equation}\label{eq:j*}
j_* = \frac{\int_0^\infty \Sigma(r) \, v(r) \, r^2 \, dr}{\int_0^\infty \Sigma(r) \, r \, dr}
\end{equation}
where $r$ is the radius, $\Sigma(r)$ is the surface mass density at radius $r$, and $v(r)$ is the rotational velocity at radius $r$. Thus one need to derive surface stellar mass density profile and rotation velocity as a function of radius, in a spatially resolved way for both, to calculate the stellar specific angular momentum.

First we evaluate the surface stellar mass density profile through multi-wavelength SED fit in a pixel-by-pixel way. Here we briefly describe the method, while details will be reported in another paper (H. Umehata et al, in preparation, see also a previous work (\cite{2023ApJ...958...36S})). We use the F115W, F200W, F356W, and F444W images, matching the PSF to that of F444W. To account for the dust-obscured star-formation activity, we also include a spatially resolved 870\,$\mu$m dust continuum map at $0.2^{\prime\prime}$ sampling.
Utilizing the F444W image as a detection mask, we use the {\sc magphys} code (\cite{2012IAUS..284..292D}) to fit the SED of each pixel. We adopt the median of the stellar mass distribution as a representative value for a pixel, while lower and upper values are determined as 16th and 84th percentiles. 

%
We use the stellar mass map to derive stellar mass density profile through elliptical aperture photometry. The  profile obtained is shown in Fig.~\ref{fig:sAM}a. There is a significant excess of stellar mass surface density at the innermost region ($R\lesssim2$\,kpc) above a simple exponential disk, which reflects the stellar bulge in ADF22.A1. Note that the panchromatic SED fit for the whole galaxy including X-ray suggests that the contribution from the AGN in the F444W photometry is limited ($\lesssim5$\%) and does not have significant impact on the derived stellar mass profile (Fig.~\ref{fig:lightning}). As some previous works (e.g., \cite{2023A&A...672A.106L}), we fit the observed profile with the combination of the bulge and disk components. For the bulge, we adopt the S\'ersic profile with $n=3$, while we assume an exponential profile for the disk. The best-fit model is shown in  Fig.~\ref{fig:sAM}a. The best-fit model is not significantly effected by S\'ersic index for a range of $n=2-4$. There are potential excess at $z\sim2-3$\,kpc between the observed and modeled profile, which may be due to a stellar bar, but the limited sampling of the profile prevent us from adding the third component. We evaluate the bulge to total stellar mass ratio $B/T\approx0.2$ from the excess over the disk exponential profile in the inner part. Since the currently available photometry taken with JWST and ALMA is not sufficient to uncover the stellar mass surface distribution in outer parts due to the lack of sensitivity, we extrapolate the profile at $R\gtrsim8$\,kpc. 


%
Until recently, stellar specific angular momentum was approximated with simplified profiles of stellar mass distribution and velocity at high redshift (\cite{2012ApJS..203...17R}). Now, we can derive the parameter with the measured stellar mass distribution, obtained from the pixel-by-pixel SED fit including submm photometry, and a rotation curve, assuming that gas kinematics trace that of the stellar component (\cite{2014ApJ...784...26O}).
To suppress the influence of non-circular motion in the bar region, we adopt the arctangent model as the rotation velocity profile. Then the stellar specific angular momentum ($j_*$) is calculated following equation (\ref{eq:j*}). The cumulative profile of stellar specific angular momentum is plotted in Fig.~\ref{fig:sAM}b, together with profiles of stellar mass distribution and rotation velocity. The total stellar specific angular momentum is calculated as a converged value, $j_*=3400\pm600$\,km\,s$^{-1}$\,kpc.

While the above method is the most direct way to derive $j_*$, it is usually quite challenging to obtain both resolved rotation curve and stellar mass surface density profile for galaxies at high redshift due to the limited sensitivity, spatial resolution, and available range of photometries. As an approximation, the scaling relation among specific angular momentum, rotation velocity and disc size and morphology has been suggested (\cite{2012ApJS..203...17R}):
\begin{equation}
    \tilde{j_*}=k_n v_s r_{\rm e}
\end{equation}
where $v_s$ is the rotation velocity at $2\times r_{\rm e}$. The remaining parameter $k_n$ is a numerical coefficient that depends on the S\'ersic index ($n$) of the galaxy:
\begin{equation}
    k_n = 1.15+0.029n+0.062n^2
\end{equation}
Assuming that we only have effective radius in an image (here F444W is adopted),
 this approximation gives $\tilde{j_*}\simeq5100$\,kpc\,km\,s$^{-1}$ which is $\sim50$\,\% larger than the estimate based on the resolved profile and rotation curve. There are several plausible factors to make the estimate of the stellar angular momentum uncertain in the approximation. First the F444W emission profile of ADF22.A1 is not perfectly modeled with a single S\'ersic profile as there is residual emissions (Fig.~\ref{fig:galfit}). ADF22.A1 has a bulge and spiral arms, and there is a limitation to fitting with a single component. Second, the stellar emission in the F444W image would not be necessarily a good tracer of stellar mass distribution. The rest-frame wavelength 11000\,\AA, which can be still affected by significant dust attenuation in massive star-forming galaxies including DSFGs. It is reported that dust attenuation can result in flatter light profiles that yield larger effective radii
 (\cite{2023ApJ...946...71C}; 
 \cite{2024arXiv240116608L}). This can also lead an overestimate of specific stellar angular momentum.
The result demonstrates that calculation of specific stellar angular momentum based on resolved profiles of stellar mass surface density and rotation velocity for galaxies at high-redshift is critically important.

\subsection{Molecular gas mass and Toomre-Q paramter} \label{analysis:toomreq}

\begin{figure}
\includegraphics[width=0.95\linewidth]{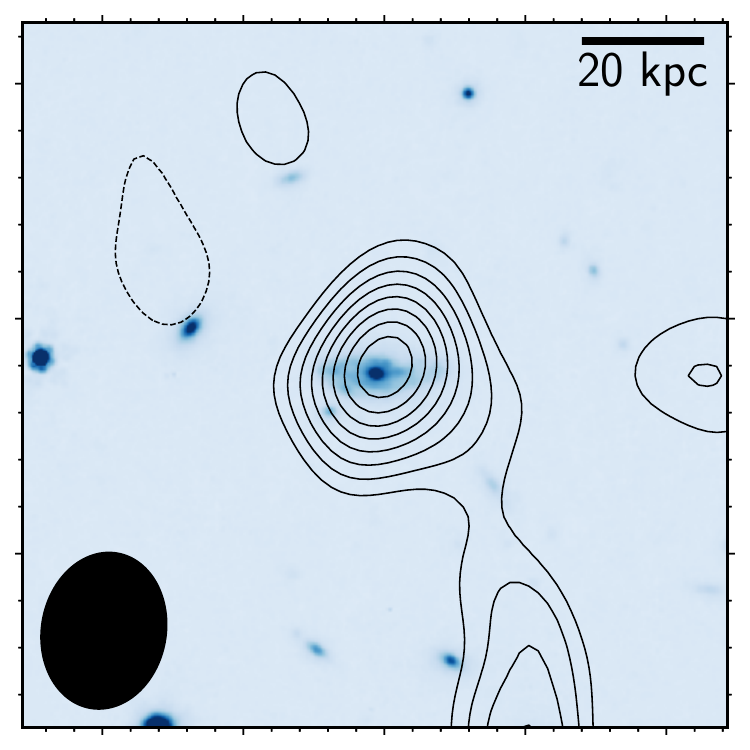}
    \caption{
    JVLA CO(1--0) intensity map of ADF22.A1. Contours are steps of $1\sigma$ starting from $\pm3\sigma$ (negative contours are dotted). The background image is the JWST/NIRCam F444W image and we show the synthesized beam in the lower left corner.
    }
    \label{fig:co10}
\end{figure}

To investigate the cause of the starburst activity seen in ADF22.A1, 
the Toomre-$Q$ parameter was computed for ADF22.A1, utilizing the resolved gas kinematics traced by [C\,{\sc ii}] and total molecular gas mass measured with CO(1--0) emission.
The total molecular gas mass of ADF22.A1 was estimated utilizing the line luminosity of the CO(1--0) emission. The ground $J = 1-0$ transition  offers a well-established tracer of the cold molecular gas
reservoirs of galaxies. The JVLA observations have successfully detected the emission in ADF22.A1, as the velcoity-integrated flux image is shown in Fig.~\ref{fig:co10}.
The total line flux measured using the  {\sc casa/imfit} task was ($0.20\pm0.03$)\,Jy\,km\,s$^{-1}$, which corresponds to a line luminosity $L^{\prime}_{\rm CO(1-0)}=(8.1\pm1.4)\times10^{10}$\,K\,km\,s$^{-1}$\,pc$^2$. 
This gives the molecular gas mass $M_{\rm gas}=(\alpha_{\rm CO}/2.5)\times(2.0\pm0.4)\times10^{11}$\,M$_\odot$. It is known that $\alpha_{\rm CO}$ vary among galaxies (for example, $\alpha_{\rm CO}\simeq0.8$ for local ULIRGs and $\alpha_{\rm CO}\simeq4.6$ for the Milky-Way like galaxies (\cite{2005ARA&A..43..677S} and references therein).
We adopt $\alpha_{\rm CO}\simeq2.5$ as a representative value, which has been derived through comparison between CO and [C\,{\sc i}] line luminosities for a set of DSFGs (\cite{2017MNRAS.466.2825B}).
The molecular gas mass measurement provides an estimate of gas mass fraction $f_{\rm gas}=M_{\rm gas}/(M_{*}+M_{\rm gas})=0.45\pm0.10$.


The Toomre-Q parameter (\cite{1964ApJ...139.1217T}) was calculated to examine the average stability of the galactic disc against local gravitational collapse. The parameter is defined as $Q=\sigma \kappa / (\pi G \Sigma_{\rm gas})$, where $G$ is the gravitational constant, and $\Sigma_{\rm gas}$ is the gas mass surface density.
Here $\sigma$ is the velocity dispersion and $\kappa$ is the epicyclic frequency. Since A1 shows a flat rotation curve, we approximate it as $\kappa=\sqrt{2}V_{\rm rot}/r$ where $V_{\rm rot}$ is the rotation velocity of the disc at radius $r$. 
The CO(1--0) emission is not well resolved with JVLA (Fig.~ \ref{fig:co10}).
Here we assumed that the CO(1--0) profile is the same as that of [C\,{\sc ii}]. We calculate the Toomre-Q parameter, calculating $\kappa$ at $r_{\rm e, [CII]}$ and averaged molecular gas mass surface density within the radius. We adopt the velocity profile and velocity dispersion measured with the [C\,{\sc ii}] modeling. The calculated $Q=0.7\pm0.2$ supports a scenario that the gas disk is gravitationally unstable and the instabilities can develop on scales larger than the Jeans length and drive active star formation observed across the disk (Fig.~\ref{fig:almamap}).

\section{Discussion}

\subsection{Origin of the rapidly rotating, giant disk}

\begin{figure} \includegraphics[width=0.95\linewidth]{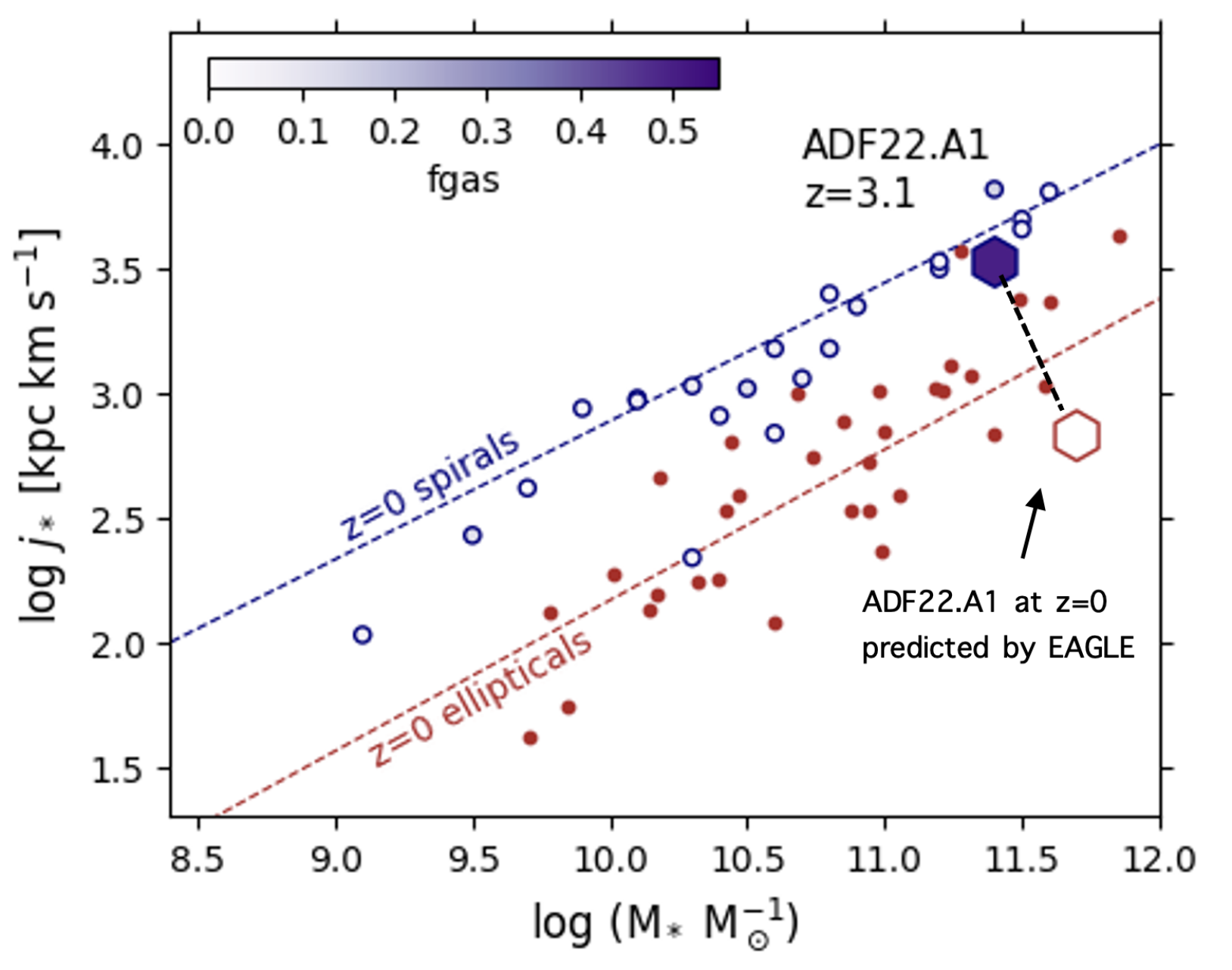}
    \caption{
    The relation between stellar mass and stellar specific angular momentum for ADF22.A1 compared to local spiral galaxies (\cite{2014ApJ...784...26O}; \cite{2018ApJ...868..133F}). ADF22.A1 has a high stellar specific angular momentum equivalent to local, massive spiral galaxies, indicating a mechanism to effectively fuel gas and angular momentum in the early universe. ADF22.A1 is expected to evolve into massive elliptical galaxies in a local cluster core, losing its angular momentum by $z\approx0$. A prediction from the EAGLE simulation illustrates the expected evolution in $j_*$
    (\cite{2016MNRAS.460.4466Z}).
    {Alt text: A graph including two lines and data points to show specific stellar angular momentum.}
    }
    \label{fig:j_ms}
\end{figure}


The derived relation between stellar mass and specific stellar angular momentum is plotted in Fig.~\ref{fig:j_ms}. For comparison, we also show the relation measured for local spiral galaxies (\cite{2014ApJ...784...26O}; \cite{2023MNRAS.518.6340D}) and ellipticals (\cite{2018ApJ...868..133F}), together with the best-fit functions (\cite{2018ApJ...868..133F}). For ADF22.A1 and local spiral galaxies, markers are color-coded with molecular gas mass fraction $f_{\rm gas} = M_{\rm mol}/(M_*+M_{\rm mol})$. 

The specific stellar angular momentum of ADF22.A1 is as high as seen in the most massive and largest local spiral galaxies (\cite{2014ApJ...784...26O}), following the local $M_*-j_* $~relation for spiral galaxies (\cite{2012ApJS..203...17R}). 
Therefore, the nature of ADF22.A1 is likely reflects a high stellar mass and a rapidly-rotating disk.
This also suggests that the association with the Hubble sequence (``the mass-spin-morphology relation'')(\cite{2013ApJ...769L..26F}), has already emerged at $z\sim3$. 
However, a distinct characteristic between ADF22.A1 at $z=3.09$ and local spirals is the molecular gas mass fraction. ADF22.A1 
 is a starburst and has an implied gas mass fraction of $\sim$45\%, which is around two orders of magnitude higher than local spirals (Fig.~\ref{fig:j_ms}). 
 There must be a source which supplies angular momentum and gas fuel to form such a system - rapidly rotating gas disk - only $\sim2$ billion years after the Big Bang.

\begin{figure}
\includegraphics[width=0.9\linewidth]{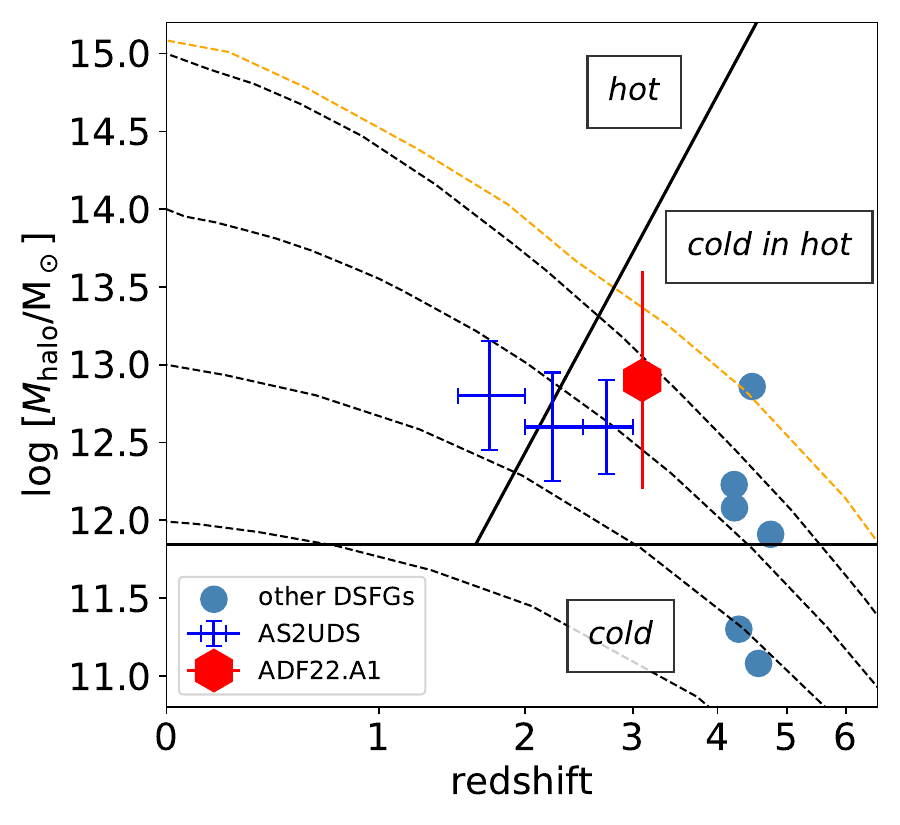}
    \caption{
    The relation between redshift and halo mass. Black dashed lines show evolution of dark matter halos, while the orange line shows the most massive halos in the $\lambda$CDM universe (\cite{2016A&ARv..24...14O}). The black solid lines separate the diagram into three regimes; ``hot'', ``cold'', and ``cold in hot'' (\cite{2006MNRAS.368....2D}). While a cold stream is not expected in the ``hot'' regime due to virial shock heating, cold stream accretion is predicted to penetrate the hot media in the most massive halos in the ``cold in hot'' regime (\cite{2016A&ARv..24...14O}). The derived halo mass of ADF22.A1 at $z=3.1$ is log($M_{\rm h}$/M$_\odot$)$=12.9\pm0.7$ and is one of the largest values among DSFGs (measured with clustering analysis (\cite{2021MNRAS.504..172S})). ADF22.A1 is located in the ``cold in hot'' regime, suggesting active cold accretion onto the galaxy regardless of the large halo mass.
    }
    \label{fig:zhalo}
\end{figure}

A plausible mechanism for forming a large, rapidly rotating disk is cold stream accretion from the cosmic web (\cite{2006MNRAS.368....2D}; \cite{2009Natur.457..451D}). The redshift-$M_{\rm h}$ diagram is proposed by \citet{2006MNRAS.368....2D} as a diagnostic of whether cold mode accretion is allowed or prohibited (Fig.~\ref{fig:zhalo}). 
Cosmological simulations show that these cold gas flows have a specific angular momentum much higher than that of the host halo and are able to quickly form into an extended disk of cold gas in 
$t\lesssim1$\,Gyr (\cite{2013ApJ...769...74S}; \cite{2022MNRAS.510.3266K}). 

We estimate dark matter halo mass ($M_{\rm h}$) hosting ADF22.A1 utilizing the empirical relation between the halo mass and flat rotation velocity for late-type galaxies (\cite{2019MNRAS.483L..98K}): 

\begin{equation}
M_{\rm h}= A {\rm log}_{10}(V_{\rm flat}/{\rm km} {\rm s}^{-1})+B 
\end{equation}

where $A=2.216\pm0.208$ and $B= 6.907\pm0.471$ for the NFW  halo profile (\cite{1996ApJ...462..563N}).
Using the flat rotation velocity derived above, the relation gives log($M_{\rm h}$/M$_\odot$)$=12.9\pm0.7$. 
In Fig.~\ref{fig:zhalo}, we plot the relation between redshift and dark matter halo mass, comparing with model tracks of the mass growth of dark matter halos and derived values in previous works for host halos of DSFGs (with clustering analysis (\cite{2021MNRAS.504..172S}) or decomposition of rotation curves (\cite{2020Natur.584..201R}; \yearcite{2021MNRAS.507.3952R}; \cite{2021Sci...371..713L}). The host halo of ADF22.A1 is one of the most massive halos known at $z\simeq3$ and also the most massive among host halos of DSFGs in the early universe.
ADF22.A1 corresponds to the ``cold in hot'' regime as marked Fig.~\ref{fig:zhalo}. This suggests that while ADF22.A1 resides in a very massive halo, cold mode accretion can still survive to provide fuel and angular momentum for the galaxy.
Intriguingly ADF22.A1 resides in the massive halo, which is associated with plentiful cool gas visible as a Mpc-scale network of Ly\,$\alpha$ filaments (\cite{2019Sci...366...97U}). Such an environment is predicted to sustain efficient cold gas accretion.

In addition to cold accretion, there is another possible contributor.
Mergers of gas-rich spiral galaxies are considered to converge to a larger, massive spiral disk satisfying the $M_*-j_* $~relation (\cite{2009ApJ...691.1168H}). Mergers may hence account for the nature of ADF22.A1, coupled to the formation of gas-rich, rapidly-rotating pre-merger galaxies from cold accretion.
The resultant giant gas disc has significant amount of molecular gas mass while rotation velocity is large. As a consequence the Toomre-Q parameter below unity as estimated in \S \ref{analysis:toomreq} and can show starburst nature due to disk instability.

\subsection{Bar-driven bulge formation} \label{dis:bar}


\begin{figure*}
\includegraphics[width=0.99\linewidth]{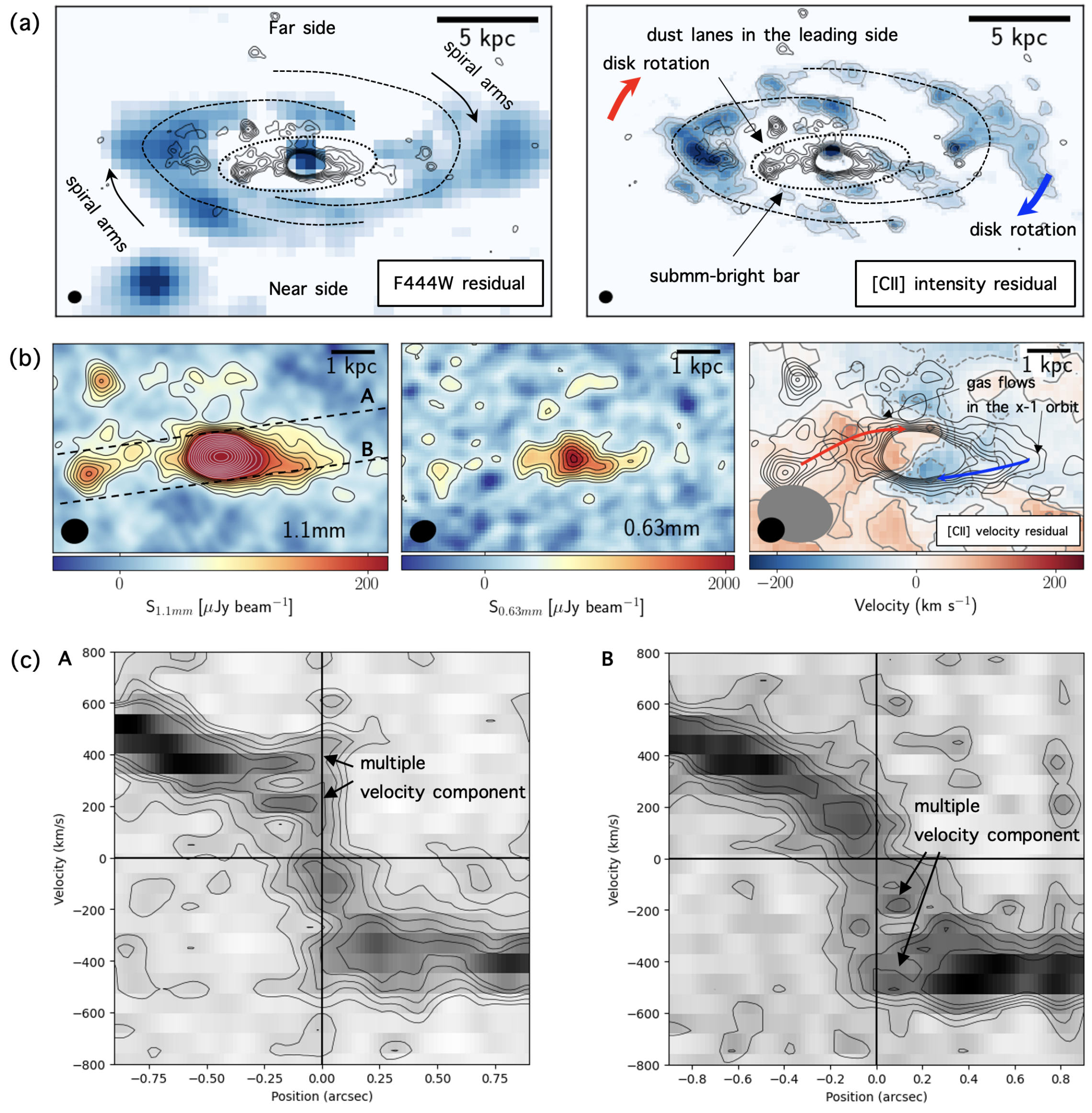}
    \caption{
    (a) The residual images of F444W and [C\,{\sc ii}] intensity maps after subtracting the best-fit models. Dashed lines and ellipticals outlines spiral arms and the bar region. Dust lanes are visible in the leading side, as are seen in local barred galaxies (\cite{1992MNRAS.259..345A}). Contours are ALMA 1.1\,mm as explained below.
    (b) The left panel shows ALMA 1.1\,mm dust continuum image at 0.07\,$^{\prime\prime}$ toward the central region. Black contours are in $1\sigma$ steps from $3\sigma$ to $10\sigma$, while white contours are in steps of $2\sigma$ from $12\sigma$. A compact dusty core and accompanied offset ridges in a bar are discovered.
    The central panel shows the case of 630\,$\mu$m dust continuum map. Contours show $2\sigma, 3\sigma, ..., 10\sigma$. The dusty core and dust lanes are detected independently at the two different wavelengths.
    The right panel shows the residual [C\,{\sc ii}] velocity field after subtraction of the best-fit model in the same field as the two left most panels. The twisted pattern coincident with the offset ridges, indicating that gas flows in the  $x-1$ orbit in the bar is likely responsible for the non-circular motion.
    (c) Two PV-diagrams along the offset-ridges (positions are shown in  panel (b) denoted as `A' and `B'). In both cases, multiple velocity components are identified along the offset ridges. Gas flows in the bar  (panel (b)) have smaller velocity than a bulge or disk (panel (a)), which causes multiple velocity components in the bar region.
    }
    \label{fig:bar_summary}
\end{figure*}

\subsubsection{a bar traced in dust continuum}

We further explored the inner regions of the galaxy where dust emission is concentrated. 
The 0.08$^{\prime\prime}$ resolution 1.1\,mm dust continuum map shows that ADF22.A1 has a bright dusty core (with the effective radius, accompanied with a pair of curved dust lanes at a $r\approx3.3$\,kpc scale.
The spatial relation between the spiral arms and the dusty core and dust lanes in ADF22.A1 is illustrated in Fig.~\ref{fig:bar_summary}a, while the structure of the dust emission is shown in  Fig.~\ref{fig:bar_summary}b. 
As illustrated, the high-fidelity 1.1\,mm dust continuum image of the inner region uncovers a pair of curved dust lanes located in the leading side, associated with the central, dusty core. The same structure is present in the 630\,$\mu$m map, demonstrating that these structures are securely detected. 
The dust lanes are called the ``offset ridges'', which form in the leading side (\cite{1992MNRAS.259..345A}). Such structures of the ISMs is a common feature seen in local barred spiral galaxies and have been reproduced by numerical and analytical calculations, including the cloud-orbit and hydrodynamical shock wave models (\cite{1980A&A....92...33C}, \cite{1991MNRAS.252..210B}, \cite{1992MNRAS.259..345A}, \cite{2023ApJ...944L..15S}). Thus the resolved structure traced by dust continuum shows that ADF22.A1 has a bar.

The existence of bars in star-forming galaxies at high redshift have been discussed in previous works (e.g., \cite{2023Natur.623..499C}; \cite{2024MNRAS.527.8941T}). Bars are suggested based on elongated morphology in dust emission (\cite{2019ApJ...876..130H}; \cite{2019MNRAS.490.4956G}),though signatures are not seen in stellar emission (\cite{2024arXiv240715846H}), while morphology of stellar emission show hints of a bar in some DSFGs (\cite{2023ApJ...958...36S}; \cite{2023ApJ...958L..26H}).
ADF22.A1 shows the most detailed map of dust emission in DSFGs, which allows us to delineate the offset ridges in the bar. 
%
Profile fit for the dusty core in the high angular resolution image (0.04$^{\prime\prime}$) shows $r_{\rm e}=670\pm30$\,pc and $n=0.6\pm0.1$.
The bar length is roughly estimated as the major axis of elliptical apertures which encompass the offset ridges delineated with 1.1\,mm dust continuum (Fig.~\ref{fig:bar_summary}a). The estimated bar size is $r\approx3.3$\,kpc.

The bar structure is not clearly visible in other tracers, including stellar light or [C\,{\sc ii}] intensity, as has also been reported for some DSFGs (\cite{2024arXiv240715846H}).
This is most likely due to severe dust attenuation. The fact that the region is bright in dust continuum indicates that dust obscuration is significant.
A similar situation, a visible dust bar which is not seen in stellar light due to high extinction is reported for the local barred spiral NGC~253 (\cite{2003AJ....125..525J}).
While the F444W image allows us to observe $\sim11000$\,\AA~at rest frame, this is insufficient to overcome the high dust extinction. Second angular resolution of the two tracers, $0.2^{\prime\prime}$ (1.5\,kpc), may not be sufficiently fine to resolve the inner structure which is visible with the $0.07^{\prime\prime}$ (500\,pc) resolution.

\subsubsection{a bar traced in gas kinematics}

The offset ridges in the bar spatially coincide with the twisted pattern of the non-circular motion in the [C\,{\sc ii}] velocity field (Fig.~\ref{fig:almamap}d). 
The right panel of Fig.~\ref{fig:bar_summary}b shows the residual velocity map at 0.2$^{\prime\prime}$ after subtracting a model derived with the {\sc 3d barolo}, together with the 1.1\,mm dust continuum contours. As shown, both the approaching and receding gas motion is detected, bracketing the central dusty core. 
To further diagnose the origin of the non-circular motion, we derive two PV-diagrams along the offset-ridges (A is for the side associated with receding gas flow, while B is for the other side connected with the approaching gas flow, Fig.~\ref{fig:bar_summary}c). As shown, it is found that there are multiple velocity components in the inner region ($r\lesssim0.4^{\prime\prime}$). In addition to a high velocity component whose velocity is equivalent to that of outer regions, a low velocity component is found in both sides. 
This is what is expected in the gas flow in the $x-1$ orbit along the bar potential toward a core (e.g., \cite{2023ApJ...944L..15S}). Gas clouds flow toward the core along the offset ridge in the rotating frame with a bar pattern speed. The line-of-sight velocity of the gas flow is governed with the bar pattern speed in the innermost region, which appears as a low velocity components in the bar region (\cite{2009PASJ...61..441H}). 

Hence as summarized in Fig.~\ref{fig:bar_summary}, there is gas inflow along the offset ridges toward the dusty core, in addition to the disk rotation.
While the moderate spatial resolution of the [C\,{\sc ii}] data allows significant blending from both components in the region, the disk rotation and bar flow has different velocities as described above, which are identified as multiple velocity components and the non-circular motion with the twisted pattern.
This finding strongly supports the theoretical expectation that bars drive gas inward from the disk, triggering central star formation and contributing to the growth of pseudobulges in disk galaxies (\cite{2004ARA&A..42..603K}). 
Cosmological simulations predict that a gas-rich, short-lived ($t\lesssim1$\,Gyr) bar forms in response to various perturbations including cold accretion from the filaments and mergers (\cite{2024MNRAS.52711095B}), which can account for the case of ADF22.A1.

\subsubsection{Contribution from AGNs}

ADF22.A1 harbors an obscured AGN at the center (\cite{2010ApJ...724.1270T}; \cite{2023ApJ...951...15M}), and AGN outflow may offer another mechanism to cause a non-circular motion. While it is challenging to determine such a contribution from AGN outflow to [C\,{\sc ii}] velocity field in ADF22.A1, we note that a typical biconical outflow which is perpendicular to the disk plane is unlikely to produce the non-circular motion.
Considering the configuration that the southern side is the near side and the north side is the far side from us, which is determined by the combination of the velocity field and a pattern of spiral arms (\cite{2023ApJ...957...48G}), such an outflow would show a blue-shifted component in the north and a red-shifted component in the south. This is opposite of what is observed. 
Furthermore, the observed non-circular motion is caused by low-velocity components, which is different from fast AGN outflows often observed in AGN-host galaxies 
 (\cite{2024MNRAS.533.4287U}).
Thus observational evidences obtained so far suggest that any outflow from the AGN is not dominating the [C\,{\sc ii}] kinematics. 

\subsection{Co-evolution in the hidden early phase}

\begin{figure}
\includegraphics[width=0.99\linewidth]{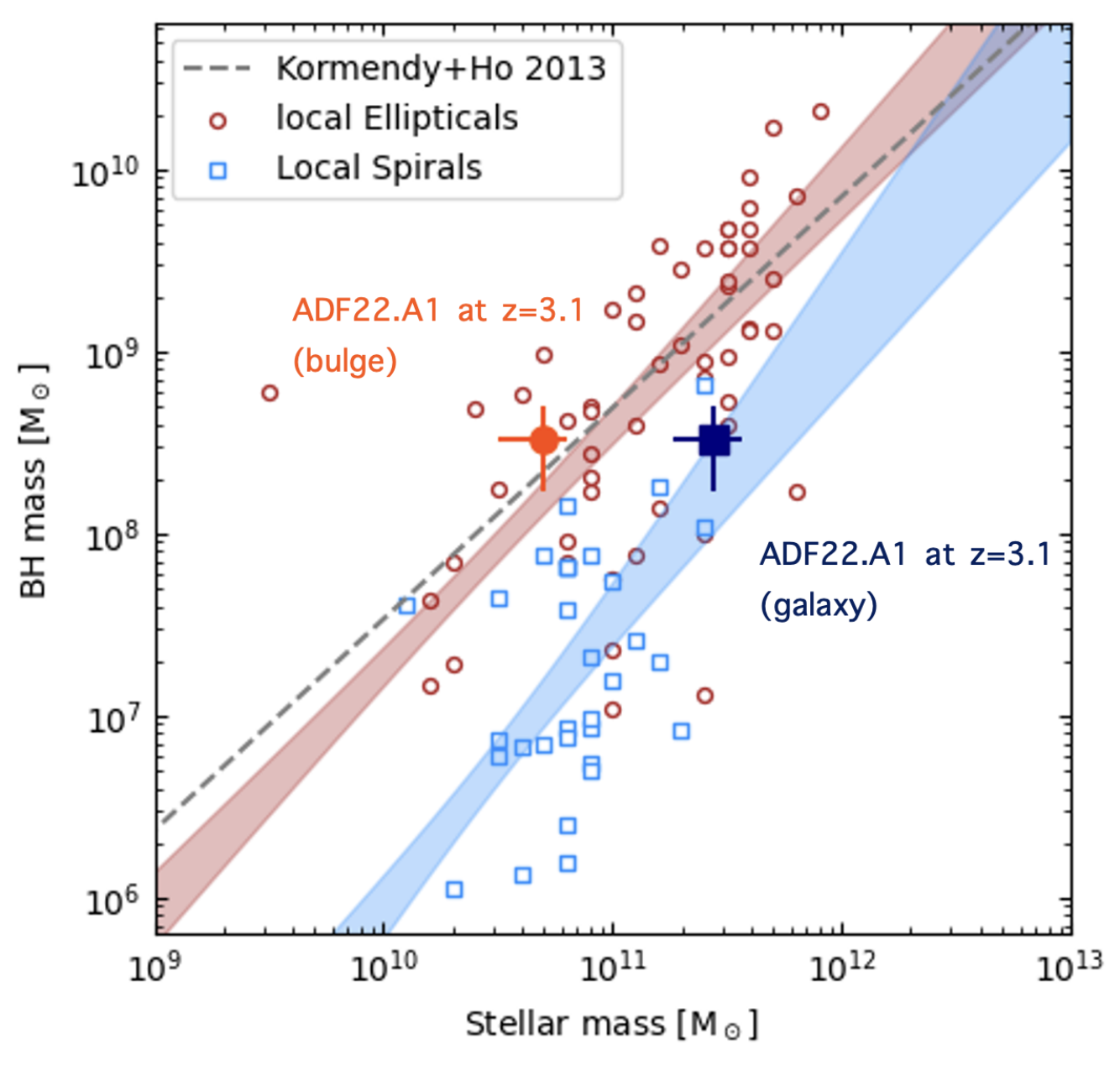}
    \caption{
    The relationship between stellar (bulge) mass and SMBH mass for local galaxies (\cite{2020ARA&A..58..257G}). Individual galaxies and best-fit functions for spiral galaxies (blue) and elliptical galaxies (red) are shown. In the case of ADF22.A1 at $z=3.1$, the relation is consistent with local spirals for the whole galaxy, while the ratio is resembles to ellipticals when comparing bulge mass.
    }
    \label{fig:smbh}
\end{figure}



%
ADF22.A1 is known to harbor a heavily obscured, but intrinsically X-ray bright (log$L_X=44.3$\,erg\,s$^{-1}$) (\cite{2010ApJ...724.1270T}; \cite{2023ApJ...951...15M}) active galactic nucleus and hence active accretion onto its central supermassive black hole is expected in the bright dusty core.
We investigated the relation between the SMBH mass and total stellar mass/bulge mass of ADF22.A1. These masses are derived based on the SED fit as described above. We plot the two ratios in Fig.~\ref{fig:smbh}. For comparison, we also show the relation for the total galaxy mass for both spiral galaxies and ellipticals in the local universe. Data for individual galaxies and best-fit functions are adopted from a recent review (\cite{2020ARA&A..58..257G} and references there in). 

As shown, the total galaxy mass-SMBH mass relation is consistent with that measured for local spirals, suggesting relatively overmassive stellar mass compared with the relation for local ellipticals.
This trend is consistent with literature which report the similar trend for DSFGs at $z\sim2$ (\cite{2008AJ....135.1968A}).
If we only focus on the bulge components, on the other hand, the mass ratio between SMBH and bulge masses $M_{\rm BH}$/$M_{\rm bulge}$ matches that of local ellipticals (\cite{2013ARA&A..51..511K}; \cite{2020ARA&A..58..257G}) (Fig.~\ref{fig:smbh}).
%
%
This suggests that the co-growth of bulges and SMBHs can  occur during the early forming phase at $z\approx3$, driven by bar structures.
This furthermore indicates that subsequent reconfiguration of the stellar mass in the current disk component of ADF22.A1 into a pressure-supported spheroid must also be associated with further growth of the SMBH.

\subsection{Evolution into local massive ellipticals}

ADF22.A1 is the most massive and active galaxy in the SSA22 proto-cluster core (\cite{2015ApJ...815L...8U}; \cite{2023ApJ...951...15M}) and a plausible analog for the progenitors of the most massive elliptical galaxies in today's dense clusters, such as Brightest Cluster Galaxies (BCGs), including cD galaxies. The results presented here give unique insights into the evolution of the most massive galaxies and SMBHs.

ADF22.A1 is a spiral galaxy although it is located in a proto-cluster core. This exhibits an inverse relation between galaxy morphology and environment compared with the local clusters (\cite{1980ApJ...236..351D}), indicating that the most massive galaxies in overdense environments must evolve from late-type rotationally dominated disks to early-type dispersion-dominated ellipticals.

To achieve this, ADF22.A1 will have to experience a significant loss of angular momentum. 
The EAGLE simulation suggests that bulge-dominated galaxies (with the bulge-to-total stellar mass ratio $B/T>0.5$) lose the majority of their specific stellar angular momentum ($80\%$ is the median value) from ``turnaround'' (at $z\sim3$) to $z\sim0$ (\cite{2016MNRAS.460.4466Z}), associated with mergers of galaxies and dynamical friction during the inner dark matter halo's assembly (\cite{2017MNRAS.464.3850L}; \cite{2018MNRAS.473.4956L}). In Fig.~\ref{fig:j_ms}b, we plot a predicted descendant at $z=0$ of ADF22.A1, assuming that (i) the molecular gas mass is all converted into stars (ii) other process (additional inflow, mergers, outflows) does not affect the stellar mass of the descendant, and (iii) 80\% of the specific stellar angular momentum is lost. As shown, the net loss predicted by the simulation sufficiently reduce the angular momentum to the levels seen in elliptical galaxies. 



\section{Conclusion}

We present JWST NIRCam imaging and ALMA imaging spectroscopy to resolve the inner structures and kinematics on (sub-)kpc scales in a bright DSFG located at the core of a $z=3.09$ proto-cluster. Our main findings include:

\begin{itemize}
 \item NIRCam images reveal a spiral-like stellar structure tracing rest-frame optical-to-near-infrared emissions. The measured effective radius, $r_{\rm e} = 7.0 \pm 0.1,\mathrm{kpc}$, is more than twice the typical size of coeval, equally massive galaxies and is comparable to that of local galaxies at $z = 0$--1, suggesting accelerated size growth in the proto-cluster core.
 \item The ALMA 870\,$\mu$m image reveals that the dust continuum is not just concentrated in a core, as seen in some DSFGs, but is distributed across the disk. This indicates that active star formation is also occurring in the disk, accompanied by significant dust production. Statistical samples with resolved studies will reveal if this is general.
 \item The [C\,{\sc ii}] velocity field is primarily dominated by rotation, with non-circular motions also detected in the center. Kinematic modeling of the [C\,{\sc ii}] emission reveals a flat rotation curve with $V_{\rm rot} = 530 \pm 10$\,km\,s$^{-1}$ out to $r\sim15$\,kpc, indicating that ADF22.A1 is an unusually fast-rotating, giant spiral galaxy.
 \item The derived $M_*$-$j_*$ relation tells us that ADF22.A1 at $z=3.09$ follows the mass-spin-morphology relation seen in the local universe. This requires a mechanism to highly spin up the disk of ADF22.A1 only $\sim$2 billion years after the Big Bang. Cold accretion is the most plausible case, coupled with mergers.
 \item The derived $M_*$-$j_*$ relation indicates that ADF22.A1 at $z=3.09$ follows the mass-spin-morphology relation observed in the local universe. This suggests that a mechanism must have rapidly spun up the disk of ADF22.A1 within only $\sim$2 billion years after the Big Bang. Cold accretion, combined with mergers, is the most plausible explanation.
 \item A bright, compact dusty core is found at the galaxy’s center, indicating the active growth phase of a proto-bulge. The estimated mass ratio of the bulge to the SMBH aligns with the local relation, while the mass ratio of the entire galaxy to the SMBH matches that of local spirals. These results suggest the emergence of bulge-SMBH co-evolution at $z \sim 3$, driven by bars.
 \item Comparison with the EAGLE simulation suggests that ADF22.A1 likely underwent significant angular momentum loss during halo assembly and associated mergers. Massive, giant spirals may represent an important early phase in the formation of BCGs and cD galaxies found in present-day cluster cores.
\end{itemize}

The data presented in this paper demonstrate the power of combining JWST and ALMA for well-resolved studies. Expanding the sample and obtaining higher-resolution observations will be crucial for deepening our understanding of the early formation phases of the most massive galaxies.

\begin{ack}
We thank the anonymous referee for constructive suggestions and comments.
We thank Hidenobu Yajima for discussions on perspectives of simulations; Takuma Izumi for discussions on the co-evolution;
Fumi Egusa and Fumiya Maeda for discussions on local analogs of barred galaxies; Tom Bakx for discussions on ALMA data reduction.
This work is based on observations made with the NASA/ESA/CSA James Webb Space Telescope. The data were obtained from the Mikulski Archive for Space Telescopes at the Space Telescope Science Institute, which is operated by the Association of Universities for Research in Astronomy, Inc., under NASA contract NAS 5-03127 for JWST. These observations are associated with program \#3547.
This paper makes use of the following ALMA data: ADS/JAO.ALMA\#2019.1.00008.S, 2021.1.00041.S, 2021.1.00071.S, 2019.1.00008.S, 2021.1.01406.S, 2022.1.00223.S. ALMA is a partnership of ESO (representing its member states), NSF (USA) and NINS (Japan), together with NRC (Canada), NSTC and ASIAA (Taiwan), and KASI (Republic of Korea), in cooperation with the Republic of Chile. The Joint ALMA Observatory is operated by ESO, AUI/NRAO and NAOJ.
The National Radio Astronomy Observatory is a facility of the National Science Foundation operated under cooperative agreement by Associated Universities, Inc. This work is based on the observations of Karl G. Jansky Very Large Array (VLA) (program ID: 16A-357; 21A-346). 
HU acknowledges support from JSPS KAKENHI Grant Numbers 20H01953, 22KK0231, 23K20240. 
KK acknowledges support from JSPS KAKENHI Grant Numbers
22H0493, 23K20035, 24H00004.
DI acknowledges support from JSPS KAKENHI Grant Number 23K20870.
This work was supported by NAOJ
ALMA Scientific Research Grant Numbers 2024-26A.
IS, AMS, DMA acknowledge STFC support (ST/X001075/1).
\end{ack}

\bibliographystyle{apj}
\bibliography{maintext.bbl}

\end{document}